\documentclass[11pt,a4paper]{article}
\usepackage{amsmath,amssymb}
\usepackage{mathrsfs,physics}
\usepackage{graphicx}
\usepackage{cite}
\usepackage{tabls}
\usepackage{color}
\usepackage{float}
\usepackage[height=8.6in,width=6.2in]{geometry}
\usepackage[font={small,it},margin=20pt,labelfont=bf]{caption}
\numberwithin{equation}{section}
\usepackage[hidelinks]{hyperref}
\usepackage[english]{babel}
\usepackage{comment}
\newcommand{\de}{\partial}
\newcommand{\eq}{\begin{equation}}
\newcommand{\en}{\end{equation}}
\newcommand{\eqa}{\begin{eqnarray}}
\newcommand{\ena}{\end{eqnarray}}

\newcommand{\cale}{\mathcal{E}}
 
\def\ii{\mathrm{i}}
\newcommand{\rme}{\mathrm{e}}
\newcommand{\vvev}[1]{\langle #1 \rangle}
\newcommand{\tmb}[1]{{\mbox{\tiny{#1}}}}

\makeatletter
\newcommand*{\letterdef@}{}
\newcommand*{\letterdef}[3]{%
	\def\letterdef@##1{\expandafter\newcommand\csname #1\endcsname{#2{##1}}}%
	\@tfor\@tempa :=#3\do{\expandafter\letterdef@\expandafter{\@tempa}}}
\makeatother
\letterdef{c#1} {\mathcal}{ABCDEFGHIJKLMNOPQRSTUVWXYZ} 
\letterdef{rm#1}{\mathrm} {dDmM} 

\newcommand{\rev}[1]{#1}

\begin{document}

\begin{center}
	{\LARGE \bf On the $T\bar T$ deformation of the compactified boson and its interpretation in  Lattice Gauge Theory. }
	
	\vspace*{15mm}

	{\Large E. Beratto${}^{\,a}$, M. Bill\`o${}^{\,b,c}$, M. Caselle${}^{\,b,c}$}

	\vspace*{8mm}
        
        ${}^a$ Universit\`a di Milano Bicocca, Dipartimento di Fisica ``G. Occhialini"\\
	Piazza della Scienza 3, I-20126 Milano, Italy
	\vskip 0.3cm
	${}^b$ Universit\`a di Torino, Dipartimento di Fisica\\
	and Arnold-Regge Center\\
	Via P. Giuria 1, I-10125 Torino, Italy
	\vskip 0.3cm
	${}^c$ 
	I.N.F.N. - sezione di Torino,\\
	Via P. Giuria 1, I-10125 Torino, Italy 
	
	\vskip 0.8cm
{	\small
		E-mail:
		\texttt{e.beratto@campus.unimib.it, caselle,billo@to.infn.it}
	}

	\vspace*{0.8cm}
\end{center}

\begin{abstract}
	We study the effective string description of space-like Polyakov loop correlators at finite temperature 
	with the goal of describing the behaviour of the space--like string tension in the vicinity of the deconfinement transition.  
	To this end we construct the partition function of the Nambu-Goto effective string theory  in presence of a compact transverse direction of length $L$ equal to the inverse temperature. 
	We then show that, under particular conditions, our result can be interpreted as the partition function of the $T\bar T$ deformation of the 2d quantum field theory describing a compactified bosonic field and that this mapping allows a deeper insight on the behaviour of the space-like observables of the theory. 
	In particular we show, by imposing that the spectrum of the model obeys the inviscid Burgers equation, that the $T\bar T$ deformations follow well defined
	trajectories in the parameter space $(\sigma,L)$ of the model, where $\sigma$ is the string tension,
	which are characterized by a constant value of the dimensionless compactification radius $\rho=L\sqrt{\sigma/2\pi}$. We discuss the potential usefulness of these results for studying the  space--like string tension of the underlying Lattice Gauge Theory and its behaviour across the deconfinement transition.
\end{abstract}


\section{Introduction}
\label{sec:intro}
An interesting open issue in Lattice Gauge Theory is to understand and model the behaviour of the so called ``space--like string tension" \cite{Karkkainen:1993ch,Bali:1993tz,Karsch:1994af,Caselle1994f,Caselle:1993cb,Koch:1994zt,Ejiri:1995gd,Sekiguchi:2016gxx,Schroder:2005zd,Liddle:2007uy,
	Cheng:2008bs,Maezawa:2007fc,Alanen:2009ej,Andreev:2006eh,Andreev:2007rx,Meyer:2005px} across the deconfinement transition. 
The space--like string tension is extracted from the correlator of space--like Polyakov loops, i.e. Polyakov loops which lay in a space--like plane, orthogonal to the compact time direction, whose size coincides with the inverse finite temperature of the theory. Due to their space--like nature these Polyakov loops  do not play the role of the order parameter of deconfinement and
the space--like \cite{Alanen:2009ej,Andreev:2006eh,Andreev:2007rx,Meyer:2005px} string tension extracted from them is different from the actual string tension of the model, which is instead extracted from time--like Polyakov loop correlators.
At low temperature the two string tensions coincide but as the temperature increases they behave differently \cite{Karkkainen:1993ch,Bali:1993tz,Karsch:1994af,Caselle1994f,Caselle:1993cb}. The ordinary string tension decreases as the deconfinement temperature is approached and vanishes at the deconfinement point, while the space--like one  remains constant and then increases in the deconfined phase \cite{Karkkainen:1993ch,Bali:1993tz,Karsch:1994af}.  The physical reason for this behavior is that the correlator of two space--like Polyakov loops describes quarks moving in a finite temperature environment. It can be shown that what we called space--like string tension is related to the screening masses in hot QCD \cite{Koch:1994zt,Ejiri:1995gd,Sekiguchi:2016gxx,Schroder:2005zd,Liddle:2007uy,Cheng:2008bs,Maezawa:2007fc} and thus it does not vanish in the deconfined phase.  

Despite the fact that it can be measured very precisely with Montecarlo simulations, a satisfactory modelization of the behaviour of the space--like string tension is still lacking.
To partially fill this gap we construct in this paper the
effective string description of the correlator of two space--like Polyakov loops assuming a Nambu-Goto form for the effective string action. This essentially amounts to extend the known effective string results to the case in which one of the transverse degree of freedom
of the model - representing the Euclidean time direction - is compact. It turns out that the natural setting to understand the behaviour of this partition function and thus model the space--like string behaviour  is  in terms of a $T\bar T$ perturbation of the free theory of a compact boson. This mapping imposes a well defined relation between the string tension $\sigma$ and the finite temperature $1/L$ of the model thus allowing to take the continuum limit in a physically meaningful way.

$T\bar T$ perturbations of 2d  Quantum Field Theories (QFT) \cite{Zamolodchikov:2004ce,Smirnov:2016lqw,Cavaglia:2016oda,Aharony:2018bad,Aharony:2018vux,Hashimoto:2019wct,Conti:2018jho,Conti:2018tca,Conti:2019dxg,Cardy:2018sdv,Caselle:2013dra,
	Dubovsky:2017cnj,Dubovsky:2018bmo,Chen:2018keo,
	Taylor:2018xcy,
	Datta:2018thy,Santilli:2018xux,Frolov:2019nrr,Cardy:2019qao,Jiang:2019hxb}
attracted a lot of interest in the past few years mainly due to the fact that they are at the
crossroad of several different research lines. 
They are related to the effective string models of the Nambu-Goto type \cite{Caselle:2013dra,Dubovsky:2017cnj,Dubovsky:2018bmo,Chen:2018keo}
and, as such, can be 
used to describe the infrared properties of extended objects such as Wilson loops or Polyakov loops correlators in the confining regime of Lattice Gauge Theories (LGTs)  \cite{Billo:2005iv, Billo:2006zg, Billo2012}.  
They are a remarkable example of an irrelevant perturbation of a 2d QFT which is integrable and whose spectrum can be easily obtained as the solution of an integrable differential equation, the inviscid Burgers equation
\cite{Zamolodchikov:2004ce,Smirnov:2016lqw,Cavaglia:2016oda,Conti:2018jho,Conti:2018tca,Conti:2019dxg}. 
They have a deep connection with gravity-like theories \cite{Dubovsky:2017cnj,Dubovsky:2018bmo,Conti:2018tca,Cardy:2018sdv}
and can be understood, in the framework of the AdS/CFT correspondence,  as a perturbation which  changes the boundary conditions of the bulk fields on $AdS_3$ (for a recent review 
see \cite{Jiang:2019hxb}).
In this paper we shall concentrate in particular on the first of these lines, looking at our effective string partition function as the $T\bar T$ perturbation of the 2d CFT of a compactified free boson.

Let us recall a few important features of this relation.
The simplest possible effective string model, compatible with the Lorentz invariance of the underlying $d$ dimensional LGT is the Nambu-Goto model 
which assumes the string action $S_{\mbox{\tiny{NG}}}$ to be proportional to the area spanned 
by the string world sheet and is parametrized by the string tension $\sigma$.  It can be shown 
that the Nambu Goto action in the physical gauge is equivalent to the $T\bar T$ perturbation of the 2d Conformal Field Theory (CFT) of $d-2$ free bosons, the transverse degrees of freedom of a string propagating in a $d$-dimensional target space \cite{Caselle:2013dra}. If we denote by 
$S_\tmb{cl}$ the ``classical'' contribution to the action, proportional to the area of the minimal string surface, and by $S_0$ 
the action of the $d-2$ free bosons representing its transverse fluctuations, one has 
\begin{align}
\label{SN}
S_\tmb{NG} = S_\tmb{cl} + S_{0}- t \int d^2\xi\, T \bar{T}~.
\end{align}
where, with this choice of sign, the perturbing parameter $t$ is related to the string tension by $\sigma=1/(2t)$ (see section \ref{sec:TTbar} for more details).

A lot of information on the spectrum of $T\bar T$ perturbed models can be obtained 
thanks to the fact that the evolution of the energy levels as a function of $t$ obeys an inviscid Burgers equation. Also many properties of the partition functions of such models are under control \cite{Aharony:2018bad,Dubovsky:2017cnj,Dubovsky:2018bmo,Datta:2018thy}. In particular, when the perturbation parameter $t$ is positive within the sign convention employed in eq. (\ref{SN})   
it can be shown that the perturbed partition function is unique and that it satisfies a differential equation originating from the Burgers equation for the energy levels  \cite{Aharony:2018bad}; this allows to determine many of its properties, both at the perturbative and at the non-perturbative level in $t$.
In this respect, the effective string theory approach, which allows to construct in an explicit way the partition function of the perturbed model for any value of the perturbing parameter, represents a perfect laboratory to test the above approaches and, indeed, another aim of our paper is also to extend the range of explicitely known partition functions to the $T\bar T$ perturbation of the CFT of a compactified boson.  This CFT is particularly interesting because it is doubly perturbed. It admits a marginal perturbation, parameterized by the compactification radius $\rho$ which moves the theory along the critical line in the $c=1$ plane \cite{Ginsparg:1988ui}, where $c$ is the central charge of the CFT. This corresponds to moving the theory along the critical low temperature phase of the $XY$ model, which is the most important statistical mechanics realization of the compactified boson universality class. There is however also the irrelevant $T\bar T$ perturbation. We shall see that the model shows a range of interesting different behaviours due to the interplay between the two perturbing parameters $\rho$ and $t$.

A final remark on the choice of the boundary conditions. 
For the LGT applications, the most natural choice of boundary conditions 
corresponds to a cylindrical world-sheet with Dirichlet boundary conditions along the Euclidean time direction -- or along a compact space-like directions in our case --  which can be immediately mapped into the correlator of two Polyakov loops and hence to the interquark potential. From the CFT point of view instead the most natural choice is the torus geometry which allows to use modular transformations
to study the nature and the location of singular  points in the perturbing parameter and to relate among them these singular behaviours.  We shall discuss both cases in the following, constructing  both the torus and the cylinder partition functions.

This paper is organized as follows. In section~\ref{sec:EST} we shall explicitely construct, both for the torus and the cylinder geometry, the Nambu-Goto effective string model  
describing the confining regime of a $d$ dimensional Lattice Gauge Theory in which one of the $d-2$ transverse directions is compactified. At the end of the section we shall discuss the LGT interpretation of our result.  In section~\ref{sec:TTbar} we discuss the interpretation of our  Effective String Theory results  as the $T\bar T$ perturbation of a compactified boson and obtain the relation imposed by integrability between the string tension $\sigma$ and the finite temperature $1/L$ of the model. Finally, section~\ref{sec:conclusion} will be devoted to a discussion of potential applications of our results and to some concluding remarks.

\section{Partition functions of  the Nambu-Goto effective string model}
\label{sec:EST}
In this section we review some aspects of the effective string description of the confining regime of gauge theories, focusing on two observables, the correlator of Polyakov loops and the interface. The first-order approach to the Nambu-Goto (NG) string allows to explicitly obtain the exact form of the corresponding partition functions \cite{Billo:2005iv,Billo:2006zg}. We work here in a compactified set-up, not yet considered in the previous literature. 

\subsection{The effective string set--up}
\label{subsec:effsetup}
\begin{figure}[hbt]
	\begin{center}
		\begingroup%
		\makeatletter%
		\providecommand\color[2][]{%
			\errmessage{(Inkscape) Color is used for the text in Inkscape, but the package 'color.sty' is not loaded}%
			\renewcommand\color[2][]{}%
		}%
		\providecommand\transparent[1]{%
			\errmessage{(Inkscape) Transparency is used (non-zero) for the text in Inkscape, but the package 'transparent.sty' is not loaded}%
			\renewcommand\transparent[1]{}%
		}%
		\providecommand\rotatebox[2]{#2}%
		\newcommand*\fsize{\dimexpr\f@size pt\relax}%
		\newcommand*\lineheight[1]{\fontsize{\fsize}{#1\fsize}\selectfont}%
		\ifx\svgwidth\undefined%
		\setlength{\unitlength}{200bp}%
		\ifx\svgscale\undefined%
		\relax%
		\else%
		\setlength{\unitlength}{\unitlength * \real{\svgscale}}%
		\fi%
		\else%
		\setlength{\unitlength}{\svgwidth}%
		\fi%
		\global\let\svgwidth\undefined%
		\global\let\svgscale\undefined%
		\makeatother%
		\begin{picture}(1,0.49195403)%
		\lineheight{1}%
		\setlength\tabcolsep{0pt}%
		\put(0,0){\includegraphics[width=\unitlength,page=1]{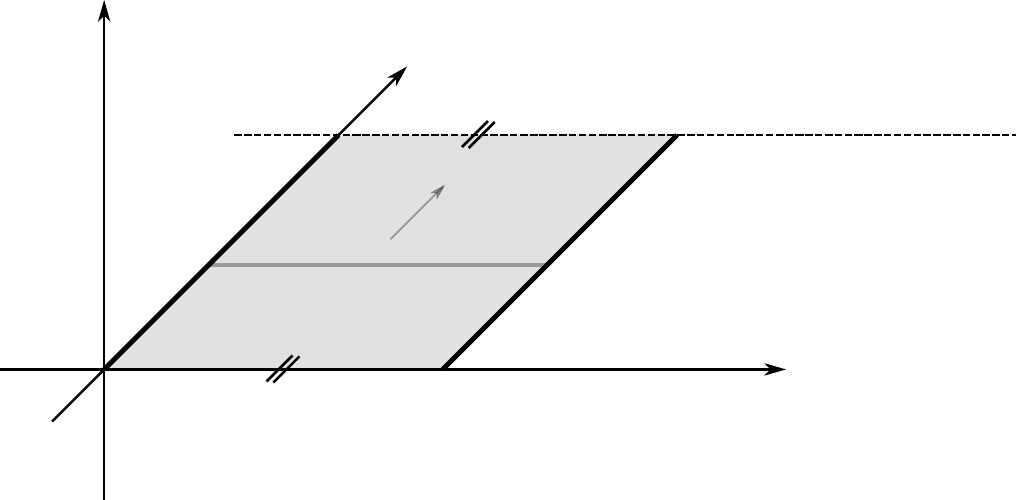}}%
		\put(0.74673705,0.08056895){\color[rgb]{0,0,0}\makebox(0,0)[lt]{\lineheight{1.25}\smash{\begin{tabular}[t]{l}$x_1$\end{tabular}}}}%
		\put(0.03675737,0.45991739){\color[rgb]{0,0,0}\makebox(0,0)[lt]{\lineheight{1.25}\smash{\begin{tabular}[t]{l}$\vec x_\perp$\end{tabular}}}}%
		\put(0.41152132,0.07692311){\color[rgb]{0,0,0}\makebox(0,0)[lt]{\lineheight{1.25}\smash{\begin{tabular}[t]{l}$R$\end{tabular}}}}%
		\put(0.23168379,0.33499606){\color[rgb]{0,0,0}\makebox(0,0)[lt]{\lineheight{1.25}\smash{\begin{tabular}[t]{l}$L_0$\end{tabular}}}}%
		\put(0.31643934,0.41757801){\color[rgb]{0,0,0}\makebox(0,0)[lt]{\lineheight{1.25}\smash{\begin{tabular}[t]{l}$x_0$\end{tabular}}}}%
		\put(0.51282053,0.29487182){\color[rgb]{0,0,0}\makebox(0,0)[lt]{\lineheight{1.25}\smash{\begin{tabular}[t]{l}$\cM_0$\end{tabular}}}}%
		\end{picture}%
		\endgroup%
		\hskip 0.2cm
		\begingroup%
		\makeatletter%
		\providecommand\color[2][]{%
			\errmessage{(Inkscape) Color is used for the text in Inkscape, but the package 'color.sty' is not loaded}%
			\renewcommand\color[2][]{}%
		}%
		\providecommand\transparent[1]{%
			\errmessage{(Inkscape) Transparency is used (non-zero) for the text in Inkscape, but the package 'transparent.sty' is not loaded}%
			\renewcommand\transparent[1]{}%
		}%
		\providecommand\rotatebox[2]{#2}%
		\newcommand*\fsize{\dimexpr\f@size pt\relax}%
		\newcommand*\lineheight[1]{\fontsize{\fsize}{#1\fsize}\selectfont}%
		\ifx\svgwidth\undefined%
		\setlength{\unitlength}{200bp}%
		\ifx\svgscale\undefined%
		\relax%
		\else%
		\setlength{\unitlength}{\unitlength * \real{\svgscale}}%
		\fi%
		\else%
		\setlength{\unitlength}{\svgwidth}%
		\fi%
		\global\let\svgwidth\undefined%
		\global\let\svgscale\undefined%
		\makeatother%
		\begin{picture}(1,0.49195403)%
		\lineheight{1}%
		\setlength\tabcolsep{0pt}%
		\put(0,0){\includegraphics[width=\unitlength,page=1]{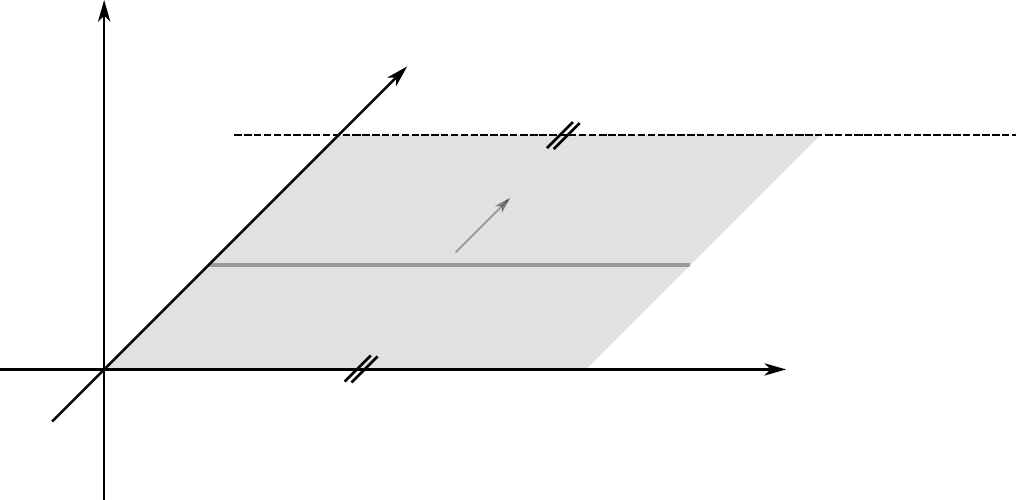}}%
		\put(0.74673705,0.08056895){\color[rgb]{0,0,0}\makebox(0,0)[lt]{\lineheight{1.25}\smash{\begin{tabular}[t]{l}$x_1$\end{tabular}}}}%
		\put(0.03675737,0.45991739){\color[rgb]{0,0,0}\makebox(0,0)[lt]{\lineheight{1.25}\smash{\begin{tabular}[t]{l}$\vec x_\perp$\end{tabular}}}}%
		\put(0.55534381,0.07692311){\color[rgb]{0,0,0}\makebox(0,0)[lt]{\lineheight{1.25}\smash{\begin{tabular}[t]{l}$L_1$\end{tabular}}}}%
		\put(0.23168379,0.33499606){\color[rgb]{0,0,0}\makebox(0,0)[lt]{\lineheight{1.25}\smash{\begin{tabular}[t]{l}$L_0$\end{tabular}}}}%
		\put(0.31643934,0.41757801){\color[rgb]{0,0,0}\makebox(0,0)[lt]{\lineheight{1.25}\smash{\begin{tabular}[t]{l}$x_0$\end{tabular}}}}%
		\put(0.60593105,0.30734199){\color[rgb]{0,0,0}\makebox(0,0)[lt]{\lineheight{1.25}\smash{\begin{tabular}[t]{l}$\cM_0$\end{tabular}}}}%
		\put(0,0){\includegraphics[width=\unitlength,page=2]{interface.pdf}}%
		\end{picture}%
		\endgroup%
	\end{center}
	\caption{Effective string description of the correlator of two Polyalov loops (left) and of an interface (right). See the text for a detailed explanation.}
	\label{fig:polint}
\end{figure}

The starting point of the effective string description of the 
inter--quark potential is to model the latter in terms of a string
partition function. One can, for instance, evaluate the inter--quark potential by using 
the expectation value $\vvev{P^\dagger (R) P(0)}$ of a pair of Polyakov loops separated by a distance $R$ in the  spatial direction $x_1$. Each Polyalov loop is the line traced in the  direction $x_0$ by an external static quark. The line is closed because this direction is compact: $x_0\sim x_0 + L_0$.
If this direction is interpreted as the Euclidean time direction, $L_0$ represents the inverse temperature of the system and we are dealing with time-like Polyakov loops. As we anticipated in the introduction, we will later take the point of view in which the 0-th direction is to be considered a spatial one. In the confining phase, the chromomagnetic flux sourced by the external quarks is squeezed into a one-dimensional string stretched between the two quarks. This open string describes, in its evolution along the temperature direction $x_0$, a surface $\cM$ with the topology of a cylinder.
In figure \ref{fig:polint}, on the left, we outlined in gray the surface $\cM$ with the minimal area, calling it $\cM_0$. We denote by 
\begin{align}
\label{AuM0}
\cA = L_0 R~,~~~ u = L_0/R
\end{align}
the area of $\cM_0$ and the ratio of its sides 

The surface $\cM$ can be described parametrically as 
\begin{align}
\label{Xmumap}
x^\mu = X^\mu(\xi^0,\xi^1)~,~~~\mu=0,1,\ldots d-1~,
\end{align}
where the adimensional parameters $\xi^\alpha$ with $\alpha=0,1$, usually referred to as world-sheet coordinates, live on a reference cylindrical surface $\Sigma$. 
The functions $X^\mu(\xi^0,\xi^1)$ represent in this language the embedding in the physical space of the world--sheet. The effective string theory approach aims at taking into account the contributions of all possible surfaces $\cM$ by path-integrating over these maps with the prescribed ``cylinder'' boundary conditions, namely fixed (Dirichlet) along the space-like directions and periodic along the temperature direction: 
\begin{align}
\label{prp0conazeff}
\vvev{P^\dagger (R) P(0)} = \int_{\tmb{cyl}} \left[ \mathcal{D} X \right] \rme^{-S_{\mbox{\tiny{eff}}}[X]} \equiv \hat\cZ_{\tmb{cyl}}(L_0,R)~,
\end{align} 
An effective string model is characterized by the choice of the string action $S_\tmb{eff}[X]$.

Let us note that the cylinder $\cM$ can also be seen, in a dual way, as the world-sheet swept out 
by a closed string emitted by a Polyakov loop (to be represented by a boundary state in the closed string Hilbert space) and reabsorbed by the other. In this set-up, it is natural to parameterise the cylinder $\cM_0$ by its area $\cA$ and the ratio $v = 1/u = R/L_0$. We will use this interpretation later.

Another observable which can be described by an effective string is the interface free energy
$F_\tmb{int}(L_0,L_1)$ of suitably chosen dual models \cite{Billo:2006zg}. This situation is depicted on the right in figure \ref{fig:polint}. Besides the direction $x_0$ (which in this context is usually interpreted as a spatial coordinate), at least one of the other directions, $x_1$ in our case, is compact with length $L_1$. In this case the surface $\cM$ has the topology of a torus, and we will have
\begin{align}
\label{inteff}
F_\tmb{int}(L_0,L_1) = \int_{\tmb{torus}} \left[ \mathcal{D} X \right] \rme^{-S_{\mbox{\tiny{eff}}}[X]} \equiv \hat\cZ_{\tmb{torus}}(L_0,L_1)~,
\end{align}
where the maps $X^\mu(\xi^0,\xi^1)$ from a reference torus $\Sigma$ to space-time describe parametrically $\cM$. The quantities 
\begin{align}
\label{AuM0torus}
\cA = L_0 L_1~,~~~ u = L_0/L_1
\end{align}
represent respectively the imaginary parts of the K\"ahler modulus (the area) and of the complex structure modulus  for the minumal surface $\cM_0$. 

\paragraph{Compact transverse directions}
In the following we shall consider the situation in which one of the transverse directions is compact, with size $L$, generalizing the results of \cite{Billo:2005iv,Billo:2006zg}. This generalization is interesting in its application to the description of Lattice Gauge theories, for the reasons described in the introduction and
discussed more in detail in section \ref{subsec:LGT}. It is interesting also within the re--interpretation of the NG string as a $T\bar T$ deformation, to be  discussed in section \ref{sec:TTbar}, in particular because of the interplay between the marginal deformation parameterized by the transverse compactification radius and the relevant $T\bar T$ deformation. Specifically, we will consider the direction $x_2$ to be compact. Note  that, even if it is not presented here, the generalization to more than one compact transverse directions is straight--forward. 

\subsection{The Nambu-Goto model}
\label{subsec:NG}
The action 
for the effective string model must be consistent with the Lorentz invariance of the underlying gauge theory. The simplest choice is the Nambu-Goto 
model, for which the string action $S_\tmb{eff}$ -- which we will call $S_\tmb{NG}$ -- is proportional to the area of $\cM$:
\begin{align}
\label{action}
S_\tmb{NG}= \sigma \int_\Sigma d^2\xi \sqrt{\det g}~,~~~
\mbox{where}~ 
g_{\alpha\beta} = \frac{\de X^\mu}{\de \xi^\alpha} \frac{\de X^\mu}{\de \xi^\beta}
\end{align} 
is the induced metric on the reference world-sheet surface $\Sigma$.
Moreover, $\sigma$ is the string tension, which has dimension $(\mbox{length})^{-2}$.

\rev{The actual effective string model which describes a gauge theory in its confining regime differs from the NG one. However, as shown in \cite{Luscher:2004ib,Aharony:2013ipa} (and in \cite{Billo:2012da} for the boundary corrections) deviations 
from the NG expression are strongly constrained by Lorentz invariance; they are expected to start at order $1/R^6$ in a large distance expansion where $R$ represents the smallest relevant length in the geometry under consideration - for instance, the Polyakov loop separation when we are in a low-temperature regime in which the length $L_0$ of the loops themselves is much bigger that $R$; for a review see for
instance \cite{Brandt:2016xsp}. In this respect the Nambu--Goto action 
can be considered as a sort of ``mean field'' approximation of the actual confining string and indeed, as expected for a mean field approximation, it gives the same answer for the confining potential of any gauge theory. Determining the higher order terms in the confining string action of a given gauge theory is a most interesting and important open issue.
One may hope to find some indications on the nature and size of the higher order corrections by studying the discrepancies  between lattice simulations of gauge theories and Nambu-Goto predictions .  
Actually, the gauge fixed Nambu-Goto action fits remarkably well the lattice data for different gauge theories, which means that deviations are rather small. In the last few years, thanks to the improvement in the precision of lattice simulations, a few signatures of these higher order corrections have been detected in the 3d gauge Ising model \cite{Caselle:2016wsw} and in the 3d SU$(N)$ gauge theories \cite{Athenodorou:2013ioa,Athenodorou:2016kpd}. The results of these papers indicate that these corrections are enhanced in the high temperature regime, i.e. when the size $L_0$ of the lattice in the direction of the Polyakov loops is slightly above the deconfinement temperature, still in the confining regime but as short as possible and represents the smallest scale in the game.} 

\rev{Let us go back to the Nambu-Goto action (\ref{action})}. 
Its reparametrization and Weyl invariances
require a gauge choice.

\paragraph{The physical gauge}
A standard choice is the so called ``physical gauge'', in which
\begin{align}
\label{physg}
x^0= X^0 = L_0\,\xi^0~,~~~ x^1 = X^1 = R\,\xi^1~. 
\end{align}
Here we are referring to the cylinder set-up, but the analogous choice can be taken also in the interface set-up. 
In this gauge, the surface $\cM$ is described by expressing the $d-2$ transverse coordinates $X^i$ with $i = 2,\ldots d-1$  as functions of the coordinates $x^0, x^1$, i.e., by giving the height profile of the surface over the minimal surface $\cM_0$ which takes over the r\^ole of the reference world-sheet $\Sigma$.   
In the physical gauge the determinant of the induced metric reduces to
\begin{align}
\label{gphysg}
\cA^{-2}\, \det g  & =
1 + \left(\frac{\de \vec X}{\de x^0}\right)^2 + \left(\frac{\de \vec X}{\de x^1}\right)^2
+ \left(\frac{\de \vec X}{\de x^0}\right)^2 \left(\frac{\de \vec X}{\de x^1}\right)^2
- \left(\frac{\de \vec X}{\de x^0}\cdot \frac{\de \vec X}{\de x^1} \right)^2~,
\end{align}
where we used the notation $\vec X= \{X^i\}$.
Inserting this into eq. (\ref{action}) we get
\begin{align}
\label{sa}
S_\tmb{NG} = S_\tmb{cl} + S_0[\vec X] + S_1[\vec X] + \ldots~,
\end{align}
where $S_\tmb{cl}=\sigma L_0 R$ is the usual ``classical'' area term, while 
$S_0[\vec X]$ is simply the action of the  two--dimensional Conformal Field Theory of $(d-2)$ free bosons living on $\cM_0$:
\begin{align}
\label{gauss}
S_0[\vec X]=\frac{\sigma}{2}\int_{\cM_0} dx^0 dx^1\, 
\frac{\de \vec X}{\de x^\alpha}\cdot \frac{\de \vec X}{\de x^\alpha}~.
\end{align}
The remaining terms in the expansion on the r.h.s. of  eq. (\ref{sa}) represent a perturbation of this free bosonic theory. In particular at the next to leading order we have
\begin{align}
\label{s1pg}
S_1[X] =  %
\sigma\int_{\cM_0} dx^0 dx^1\,\left[\frac{1}{8}
\left(\frac{\de \vec X}{\de x^\alpha} \cdot \frac{\de \vec X}{\de x^\alpha}\right)^2
- \frac{1}{2}
\left(\frac{\de \vec X}{\de x^\alpha} \cdot \frac{\de \vec X}{\de x^\beta}\right)
\left(\frac{\de \vec X}{\de x^\alpha} \cdot \frac{\de \vec X}{\de x^\beta}\right)
\right]~.
\end{align}
Actually, as it is well known \cite{Caselle2013b}, the full NG action can be seen as a $T\bar T$ deformation of the free bosonic theory. In this perspective, eq. (\ref{s1pg}) encodes just the first order in this perturbation; however, it already fixes the relation between the parameter, usually called $t$, of the $T\bar T$ perturbation and the string tension $\sigma$. This will be discussed in section \ref{sec:TTbar}. 

If we rescale the fields by setting $\vec X = \vec{\phi}/\sqrt{\sigma}$, the Nambu-Goto action takes the form
\begin{align}
\label{action2}
S_\tmb{NG}= \sigma \cA \cdot \int_\Sigma d^2\xi \sqrt{ 1+\frac{1}{\sigma \cA}
	\left(	
	\frac{\de \vec \phi}{\de \xi^\alpha} \cdot \frac{\de \vec \phi}{\de \xi^\alpha}
	\right) }~.
\end{align}
This expression
makes it explicit that the above expansion is actually an expansion in powers of 
$1/(\sigma\cA) = 1/(\sigma L_0 R)$. In particular, a part from the classical area term $S_\tmb{cl}$, the free gaussian action is recovered in the $\sigma L_0 R \to \infty$ limit . This is exactly the long string limit in the Lattice Gauge Theory context in which, as it is well known, the effective string contribution reduces to the 
so called ``Luscher term" \cite{Luscher:1980fr,Luscher:1980ac} that corresponds to  the leading term of the partition function of $(d-2)$ non-interacting bosons.

\subsection{First-order formulation}
\label{subsec:fof}
A remarkable feature of the Nambu-Goto model is that the functional integral on the transverse degrees of freedom $\vec X(\xi^0,\xi^1)$ can be 
performed explicitely using the Polyakov trick  for any choice of the boundary conditions,  see \cite{Billo:2005iv, Billo:2006zg, Billo2012} for the application to the present context (see also \cite{Luscher:2004ib} for a derivation of the NG partition function using a different approach). One starts from a first-order
reformulation of the NG model, in which the action is 
\begin{align}
\label{S1storder}
S = \frac{\sigma}{2} \int_{\Sigma}d^2\xi\sqrt{\det h}\, h^{\alpha\beta} \de_\alpha X^\mu \de_\beta  X_\mu~, 
\end{align} 
with $\mu=0,\ldots d-1$.
If one integrates out the independent world-sheet metric $h$, its e.o.m. identifies it with the induced metric $g$ of eq. (\ref{action}) and one retrieves the NG model. For each fixed topology of the world-sheet, however, one can use the re--parametrization and Weyl invariance of eq. (\ref{S1storder}) to bring $h_{\alpha\beta}$ to a reference form $\rme^\phi \hat h_{\alpha\beta}$.
The scale factor $\rme^\phi$ decouples at the classical level; at the quantum level, due to an anomaly, this is true only in the critical dimension $d=26$. However, the effect of
the anomaly is known to become irrelevant at large physical distances, for instance, for $R$ large in the cylinder case \cite{Olesen:1985pv}.  
\rev{In principle its effect should be captured  by adding to the NG action suitable higher order terms in a large distance expansion. Such kind of corrections of the same type of those that, as we remarked in section \ref{subsec:NG}, are expected to differentiate the effective string models for specific gauge theories from the NG model. For this reason our results, based on the first-order Polyakov reformulation of the NG model should be only considered as a large distance effective description of the actual confining string.} 

If $\Sigma$ is a cylinder or a torus, we can choose the so-called conformal gauge, fixing  $\hat h_{\alpha\beta} = \delta_{\alpha\beta}$ and realizing $\Sigma$ as a rectangle with one or two couples of opposite sides identified.
The action of the model reduces then to the free action describing the CFT of $d$ bosons plus the action $S_\tmb{gh}$ for the ghost-antighost system that arise from the Jacobian of the gauge-fixing procedure:
\begin{align}
\label{SCFTdgh}
S = \frac{\sigma}{2} \int_{\Sigma}d^2\xi\, \de^\alpha X^\mu \de_\alpha\beta X_\mu
+ S_\tmb{gh}~.  
. 
\end{align} 
In both the cylinder case and the torus case there is a residual Teichm\"uller parameter which we cannot change by means of conformal rescalings and which has to be integrated over. This integration is the remnant of the path-integral over the independent world--sheet metric $h$.  

We will now consider separetely in detail these two cases.

\subsection{The interface case (torus geometry)}
\label{subapp:int}
Let us now start from the case in which the world--sheet $\Sigma$ is a torus, as appropriate for the one-loop partition function of closed strings. The closed string world-sheet is periodic; in our conventions, 
$(\xi^0,\xi^1)\sim (\xi^0,\xi^1 + \pi)$.  
The complex structure of the world-sheet, $\tau=\tau_1+\ii\tau_2$ with $\tau_2\geq 0$ defines a secod identification: $(\xi^0,\xi^1)\sim (\xi^0 + \pi\tau_1,\xi^1 + \pi\tau_2)$. It
represents a Teichm\"uller parameter and has to be integrated over. In fact, the string partition function has the form
\begin{align}
\label{pftorus}
\cZ_\tmb{torus} = \int_{\mathscr{F}}  \frac{d^2\tau}{4\tau_2} Z^{(d)}(\tau,\bar{\tau}) Z_\tmb{gh}(\tau,\bar{\tau})
= \int_{\mathscr{F}}  \frac{d^2\tau}{4\tau_2^2} Z^{(d)}(\tau,\bar{\tau}) \,\tau_2 Z_\tmb{gh}(\tau,\bar{\tau})
~.
\end{align}
where $\mathscr{F}$ is the fundamental region of the Teichm\"uller space with respect to the action of the modular group $\mathrm{PSL}(2,\mathbb{Z})$ that maps $\tau$ to $\tau^\prime$, with
\begin{align}
\label{modaction}
\tau^\prime  = \frac{a\tau + b}{c\tau + d}~,~~~a,b,c,d\in\mathbb{Z}~,~~~ad-bc=1~.
\end{align}
The parameter $\tau^\prime$ describes an equivalent torus. The integral in eq. (\ref{pftorus}) is limited to the fundamental cell $\mathscr{F}$ to avoid redundancy, but this is consistent only if the integrand is modular invariant. In the second step in eq. (\ref{pftorus}) we have singled out the measure $d^2\tau/\tau_2^2$ which, according to eq. (\ref{modaction}), is modular invariant by itself. 

Moreover, $Z^{(d)}(\tau,\bar{\tau})$ is the CFT partition function of $d$ bosonic fields $X^\mu$, and $Z^\tmb{gh}(\tau,\bar{\tau})$ that of the ghost system, both computed on a fixed world-sheet of complex structure $\tau$.   

Let us describe the CFT partition functions appearing in (\ref{pftorus}), taking into account
that in our set-up three of the target space coordinates are compact:
\begin{align}
\label{periodxa}
x^a \sim x^a + L^a~,~~~a=0,1,2~,
\end{align}
while the remaining $d-3$ ones we take for simplicity to be non-compact (but it is straightforward to compactify some of them). 

\paragraph{Compact boson} The partition function for a boson field $X$ with compactification length $L$ can be easily derived in an operatorial formalism, in which the coordinate $\xi^0$ on $\Sigma$ plays the r\^ole of (Euclidean) world-sheet time. 
Since the world-sheet coordinate $\xi^1$ is periodic, the free field $X(\xi^0,\xi^1)$
can have non-trivial winding $w\in\mathbb{Z}$ defined by 
\begin{equation}
\label{eq:A.2}
X(\xi^0, \xi^1 + \pi) = X(\xi^0, \xi^1) +  w L~. 
\end{equation}
In the $w$-th winding sector, its expansion comprises left-moving and right-moving oscillators plus zero modes:
\begin{equation}
\label{eq:A.4}
X(\xi^0, \xi^1) = \hat x + \frac{\hat p}{\pi\sigma} \xi^0 + \frac{w L}{\pi } \xi^1 + \frac{\ii}{\sqrt{4\pi\sigma}} 
\sum_{k \neq 0} \left(\frac{\alpha_k}{k} \rme^{-2\ii k\xi} + \frac{ \bar{\alpha}_k^i}{k} \rme^{-2\ii k\bar\xi} \right)~,
\end{equation}
where we introduced $\xi = \xi^0 + \ii \xi^1$ and $\bar \xi = \xi^0 - \ii \xi^1$.  
Since the target space is compact, the spectrum of the momentum operator $\hat p$ is quantized: $p = 2\pi n/L$, with $n\in\mathbb{Z}$.    

The partition function is given by
\begin{equation}
\label{eq:A.6}
Z(\tau,\bar{\tau};L) = \Tr \left( q^{\mathscr{L}_0 -\frac{1}{24}} \bar{q}^{\bar{\mathscr{L}}_0 - \frac{1}{24}} \right)~,
\end{equation}
where $q = \exp(2\pi\ii\tau)$ while the $0$-th  Virasoro operators $\mathscr{L}_0$ and $\bar{\mathscr{L}}_0$ are the zero modes of the holomorphic (left-moving) and anti-holomorphic stress-energy tensor. One has:
\begin{equation}
\label{eq:A.7}
\mathscr{L}_0 = \frac{1}{8\pi\sigma} \left(\frac{2\pi n}{L} + w \sigma L\right)^2 +  \sum_{k  = 1}^\infty N_k~,~~~
\bar{\mathscr{L}}_0 = \frac{1}{8\pi\sigma} \left(\frac{2\pi n}{L} - w \sigma L\right)^2 +  \sum_{k  = 1}^\infty \bar N_k~,~~~
\end{equation}
where $N_k$ and $\bar N_k$ are the number operators for the $k$-th left-moving and right-moving oscillator systems. 
These Virasoro operators are simply related to the Hamiltonian and the
angular momentum by $H = \mathscr{L}_0 + \bar{\mathscr{L}}_0 - 1/12$ and  
$J = \mathscr{L}_0 - \bar{\mathscr{L}}_0$. The evaluation of the trace is straight--forward, and the result is
\begin{equation} 
\label{CFT2s}
Z(\tau,\bar{\tau}; L) = \frac{ 1 }{\eta(q)}\frac{1}{\eta( \bar{q})} \,
\Gamma(\tau,\bar{\tau};L)~.
\end{equation}
The factor $1/\eta(q)\eta(\bar q)$, with the Dedekind eta-function being defined in eq. (\ref{defeta}), arises from the trace over the Fock spaces of the oscillator non-zero modes. The other factor comes from the sum over the zero-mode quantum numbers and reads
\begin{align}
\label{Gammazm}
\Gamma(\tau,\bar{\tau};L) = 
\sum_{n, w \in \mathbb{Z}} 
q^{\frac{1}{8\pi\sigma} \left(\frac{2\pi n}{L} + w \sigma L\right)^2 } 
\bar{q}^{\frac{1}{8\pi\sigma} \left(\frac{2\pi n}{L} - w \sigma L\right)^2 }
=
\sqrt{\frac{\sigma}{2\pi\tau_2}}L
\sum_{m, w \in \mathbb{Z}} \rme^{- \frac{\sigma L^2}{2\tau_2}
	\left|m - \tau w \right|^2}~,  	
\end{align}    
where in the second step we used the Poisson resummation formula (\ref{poisson}).
The sum over $m,w$ represents the sum over classical solutions of the field X which, beside eq. (\ref{eq:A.2}), also have a non-trivial wrapping along the compact propagation direction:
\begin{align}
\label{wrapxi0}
X(\xi^0 + 2\pi\tau_2,\xi^1 + 2\pi\tau_1) = X(\xi^0,\xi^1) + m L~.
\end{align}   

The partition function $Z(\tau,\bar{\tau};L)$ is invariant under the modular transformations
(\ref{modaction}). This is evident using the second expression of $\Gamma(\tau,\bar{\tau};L)$ in eq. (\ref{Gammazm}), taking into account that $\sqrt{\tau_2}\eta(q)\eta(\bar q)$ is modular invariant, see eq. (\ref{Seta}), and so is the sum over $m$ and $w$, in which the effect of the 
modular transformation (\ref{modaction}) is to replace $m,w$ with 
\begin{align}
\label{modularmw}
m^\prime = d m + b w~,~~~ w^\prime = a w + c m~.
\end{align}

$Z(\tau,\bar{\tau};L)$  also displays the so-called $T$-duality, namely it is invariant under
\begin{align}
\label{Tduality}
L \to \frac{2\pi}{\sigma L}~,
\end{align}
as it is clear from the structure of the sum in the first expression of $\Gamma$ in eq. (\ref{Gammazm}). 

\paragraph{Non-compact boson} In this case the field $X$ has no winding, so we have to set $w=0$ in the expansion (\ref{eq:A.4}) and in the expression of the Virasoro operators, eq. (\ref{eq:A.7}). Moreover the momentum eigenvalue $p$ is not quantized and, in computing the partition function trace, the sum over $n$ is replaced by a gaussian integration over $p$. The final result, proportional to the regularized volume $V$ of the target space for $X$, is 
\begin{align}
\label{Znc}
Z_\tmb{n.c.}(\tau,\bar{\tau}) = \frac{ 1 }{\eta(q)}\frac{1}{\eta( \bar{q})} \,
\sqrt{\frac{\sigma}{2\pi\tau_2}} V
\end{align}
and is modular invariant. 

\paragraph{The ghost partition function}
As is well know, the partition function for the ghost system exactly cancels the non-zero mode contributions of two bosonic fields:
\begin{align}
\label{Zgh}
Z_\tmb{gh}(\tau,\bar{\tau}) = \eta^2(q) \eta^2( \bar{q})~. 
\end{align}
The combination $\tau_2 Z_\tmb{gh}(\tau,\bar{\tau})$ is therefore modular invariant. 

\paragraph{The string partition function}
Let us go back to eq. (\ref{pftorus}). Taking into account eq. (\ref{periodxa}) and eq. (\ref{CFT2s}), we have
\begin{align}
\label{Zdis}
Z^{(d)}(\tau,\bar{\tau}) = 
Z(\tau,\bar{\tau};L_0)\, Z(\tau,\bar{\tau};L_1)\, Z(\tau,\bar{\tau};L)\,
\left[Z_\tmb{n.c.}(\tau,\bar{\tau})\right]^{d-3}~.,	 
\end{align}
where we denoted the compactification radius in the direction $x^2$ by $L$.
Using this expression and eq. (\ref{Zgh}) we have the explicit integral expression of the full string partition function $\cZ_{\tmb{torus}}$ of eq. (\ref{pftorus}), and we see that the integrand is indeed modular invariant. However, to recover in our first-order formulation the quantity $\hat\cZ_\tmb{torus}$ that gives the effective string description of the interface free energy, see eq. (\ref{inteff}) and figure \ref{fig:polint}, we have to extract the contributions of the embeddings in which the world-sheet $\Sigma$ covers once the torus target space of sides $L_0$, $L_1$. 

To do so, one can consider the contribution of the zero-modes in the directions $0$ and $1$ and group the terms corresponding to the possible values of $m^a$ and $w^a$, with $a=0,1$, into orbits of the modular group. Following \cite{Dixon:1990pc}, we restrict to specific representatives in each orbits, while correspondingly enlarging the $\tau$ integration region. At this point, we extract the contributions mentioned above by  
taking $m^0=1, w^0=0, w^1=1$ and summing over $m^1$ or, equivalently, undoing the Poisson resummation, over $n^1$. The details are given in appendix \ref{app:A}. With this choice, we obtain 
\begin{align}
\label{Zhatres}
\hat\cZ_{\tmb{torus}} 
& = \frac{V_{d-3} L_0}{4} \left(\frac{\sigma}{2\pi}\right)^\frac{d-2}{2}
\int_0^\infty \frac{d\tau_2}{(\tau_2)^{1+\frac{d-2}{2}}}
\rme^{-\frac{\sigma L^2_0}{2\tau_2}}
\int_{- 1/2}^{1/2} d\tau_1 	
\left(\frac{1}{\eta(q)\eta(\bar q)}\right)^{d-2} \nonumber \\
&\times 
\sum_{ n_0,n_2,w_2 \in \mathbb{Z} } 
q^{\frac{1}{8\pi\sigma} \left[
	\left(\frac{2\pi n_1}{L_1} + \sigma L_1 \right)^2
	+ \left(\frac{2\pi n_2}{L} + w_2 \sigma L \right)^2\right]}
{\bar q}^{\frac{1}{8\pi\sigma} \left[
	\left(\frac{2\pi n_1}{L_1} - \sigma L_1 \right)^2
	+ \left(\frac{2\pi n_2}{L} - w_2 \sigma L \right)^2\right]}~,
\end{align}
with $V_{d-3}$ being the volume of the non-compact target space directions. 
Expanding in series the Dedekind eta functions according to eq. (\ref{expeta}) we obtain 
\begin{align}
\label{eq:A.16}
\hat{\mathcal{Z}}_\tmb{torus}  
& = \frac{V_{d-3} L_0}{4} \left(\frac{\sigma}{2\pi}\right)^\frac{d-2}{2} 
\sum_{k, k'  = 0}^\infty c_k c_{ k' } 
\sum_{n_1,n_2, w_2 \in \mathbb{Z}}
\int_{-1/2}^{1/2} d\tau_1 \rme^{2\pi \ii\tau_1 (k - k' + n_1 + n_2 w_2 )}
\nonumber\\ 
& \times \int_{0}^\infty \frac{d\tau_2}{\tau_2^{1 + \frac{d-2}{2}}}
\times \rme^{-\tau_2 \left(  \frac{\sigma L^2_1}{2} + \frac{\sigma L^2 w^2_2}{2} + \frac{2\pi^2 n^2_1}{\sigma L^2_1} + \frac{2\pi^2 n^2_2}{\sigma L^2}  + 2\pi (k + k' - \frac{d-2}{12} ) \right) -\frac{1}{\tau_2} \frac{\sigma L^2_0}{2}}~. 
\end{align}
The integration over $\tau_1$ produces the Kronecker delta function 
$\delta_{k - k' + n_1 + n_2m_2 }$, while the integration over $\tau_2$ can be carried out by means of the formula in eq. (\ref{eq:A.16.2}).
The final result is
\begin{equation}
\label{eqA}
\hat\cZ_\tmb{torus} 
= \frac{V_{d-3} L_0}{2}
\left( \frac{\sigma}{2\pi}\right)^{\frac{d-2}{2}}  
\sum_{k, k^\prime = 0}^\infty c_k c_{k^\prime}\sum_{n_1, n_2, w_2 \in \mathbb{Z}}
\delta_{k-k^\prime + n_1 + n_2 w_2}  
\left(\frac{\mathcal{E}}{u} \right)^{\frac{d-2}{2}} 
K_{\frac{d-2}{2}} \left(\sigma \mathcal{A}\mathcal{E} \right)~,
\end{equation}
with
\begin{equation} 
\label{eqB}
\mathcal{E} = \sqrt{1 + \frac{4\pi u}{\sigma \cA} (k + k' - \frac{d-2}{12} ) + \frac{4\pi^2 u^2 n^2_1}{(\sigma \cA)^2} + \frac{4\pi^2 u n^2_2}{\sigma^2 \cA L^2} + \frac{u L^2 w^2_2}{\cA} }~. 
\end{equation}

Considering the $L_0\to \infty$ limit  of eq. (\ref{eqA}), we extract from the asymptotic behaviour of the Bessel function
the energy spectrum. Writing $\exp(-\sigma\cA\cE) = \exp(-L_0 E)$, and remembering that $u/\cA = 1/L_1^2$ is independent of $L_0$, we find
\begin{align}
\label{eq:1.42}
E = \sigma L_1 \cE~,
\end{align}
which represents the mass of a closed string state with left- and right-moving occupation numbers
$k$ and $k'$ and quantum numbers $n_1,n_2,w_2$ subject to the level-matching condition
$k-k' + n_1 + n_2 w_2=0$. 

As a consistency check, let us consider the limit $L\to\infty$ in which all transverse directions are non compact. Following the derivation in Appendix \ref{app:A} one sees that in this limit only the trivial winding $w_2=0$ contributes, so that the $\delta$ function reduces to imposing  $n_0=k'-k$. Moreover, the KK sum over the discrete momentum $n_2$ is replaced by a gaussian integral, whose result modifies the final integration over the Teichm\"uller modulus $\tau$. In this way, one recovers the result of 
\cite{Billo:2006zg}:
\begin{equation}
\label{torus2}
\hat\cZ_\tmb{torus} = \frac{V_{d-2} L_0}{2} \left(\frac{\sigma}{2\pi}\right)^{\frac{d-1}{2}}
\sum_{k,k'=0}^\infty  c_k c_{k'}\, \left(\frac{\cE}{u}\right)^{\frac{d-1}{2}}\,
K_{\frac{d-1}{2}}\left(\sigma \cA\cE\right)~,
\end{equation}
where $V_{d-2}$ denotes the transverse volume and 
\begin{equation}
\label{torus3}
\cale
= \sqrt{1 + \frac{4\pi\, u}{\sigma \mathcal{A}}(k+k'-\frac{d-2}{12}) + 
	\frac{4\pi^{2} \, u^2\, (k-k')^{2}}{(\sigma\mathcal{A})^2}}~.
\end{equation}

As mentioned above, the partition function (\ref{eqA}) enjoys the $T$-duality symmetry
\begin{align}
\label{Td2}
L \to \frac{2\pi}{\sigma L}~.
\end{align}
It would be interesting to explore in Lattice simulations%
\footnote{Notice that the LGT realization of the torus topology is rather non-trivial and is possible only using duality.  For instance the simplest possible realization is in the 3d gauge Ising model and is given by the free energy of interfaces in the dual Ising spin model \cite{Caselle:1994df}. More complex realizations require for instance the study of the so called 't Hooft loops in non abelian LGTs \cite{deForcrand:2000fi}.} the consequences of this duality. For instance it implies the existence of a ``minimal'' compactification radius given by the self--dual point
\begin{align}
\label{L2c}
L_{\tmb{c}}=\sqrt{\frac{2\pi}{\sigma}}~.
\end{align}

\subsection{Polyalov loop set-up}
\label{subapp:pol}
Again, to compute the string partition function we use the first order set--up.
The world-sheet $\Sigma$ is a cylinder, and corresponds to the one-loop partition function of open strings. 
In our conventions, the range of the space-like world--sheet coordinate is $\xi^1\in[0,\pi]$. 
The length of the cylinder $\Sigma$ is a real Teichm\"uller parameter, which we call $t$. We have to integrate over this Teichm\"uller parameter, and the string partition function is
\begin{align}
\label{spfcyl}
\cZ_\tmb{cyl} = \int_0^\infty \frac{dt}{2t} Z^{(d)}(t) Z_\tmb{gh}(t)~,
\end{align}
where $Z^{(d)}(t)$ is the CFT cylinder partition function of $d$ bosonic fields $X^\mu$ and 
$Z_\tmb{gh}(t)$ that of the ghost system, both computed on a fixed cylinder of parameter $t$. 

\paragraph{Open string channel}
As depicted in figure \ref{fig:polint}, the open string is attached to the two Polyakov loops, which span the direction $x^0$. This means that the embedding field $X^0(\xi^0,\xi^1)$ has free (Neumann-Neumann) boundary conditions at both endpoints, while $X^1(\xi^0,\xi^1)$ has fixed (Dirichlet-Dirichlet) boundary conditions, with one end--point fixed in $0$ and the other one in $R$. The field $X^2(\xi^0,\xi^1)$ has DD boundary conditions, but can wind $w_2$ times around the compact coordinate $x^2$. The fields  $X^i(\xi^0,\xi^1)$, for $i=3,\ldots,d$ are DD, with both endpoints in $0$. Their mode expansion is therefore the following: 
\begin{align}
\label{eq:A.21}
X^0(\xi^0,\xi^1) & = \hat x^0 + \frac{\hat p_0}{\sigma\pi}\xi^0 + \frac{\ii}{\sqrt{4\pi\sigma}}  
\sum_{k \neq 0} \frac{\alpha_k^0}{k}\rme^{-\ii k\xi^0}\cos{k\xi^1}~,\nonumber\\ 
X^1(\xi^0,\xi^1) & = \frac{R}{\pi}\xi^1 - \frac{1}{\sqrt{4\pi\sigma}}  
\sum_{k \neq 0} \frac{\alpha_k^1}{k}\rme^{-\ii k\xi^0}\cos{k\xi^1}~,\nonumber\\ 
X^2(\xi^0,\xi^1) & = \frac{w_2 L}{\pi}\xi^1 - \frac{1}{\sqrt{4\pi\sigma}}   \sum_{k \neq 0} \frac{\alpha_k^2}{k}\rme^{-\ii k\xi^0}\cos{k\xi^1}~,\nonumber\\ 
X^i(\xi^0,\xi^1) & =  - \frac{1}{\sqrt{4\pi\sigma}}   \sum_{k \neq 0} \frac{\alpha_k^i}{k}\rme^{-\ii k\xi^0}\cos{k\xi^1}~,~~~i=3,\ldots, d-1~. 
\end{align}
Note that the fields with DD b.c.'s do not possess zero modes. Since the target space direction $x^0$ is compact with length $L_0$, the spectrum of the momentum operator $\hat p_0$ is discrete:
$p_0 = 2\pi n_0/L_0$, with $n_0\in\mathbb{Z}$. The cylinder CFT partition function for these fields is given by
\begin{equation}
\label{eq:A.23}
Z^{(d)}(t) = \Tr\left(q^{\mathscr{L}_0 - \frac{d}{24}}\right)~,
\end{equation}
where we $q= \exp(-2\pi t)$ and the Virasoro zero-mode operator reads
\begin{equation}
\label{eq:A.22}
\mathscr{L}_0 =   \frac{2\pi}{\sigma} \frac{n_0^2}{L_0^2} 
+ \frac{\sigma R^2 }{2\pi} + \frac{\sigma L^2 w_2^2}{2\pi} 
+ N^{(d)}~.
\end{equation}
Here $N^{d}$ is the total number operator for all the non-zero mode oscillators of the $d$ bosonic fields. The trace in eq. (\ref{eq:A.23}) receives contributions from the non-zero mode oscillators and from the zero-mode sector and is found to be given by 
\begin{align}
\label{eq:A.24}
Z^{(d)}(t) 
& = \left(\frac{1}{\eta(q)}\right)^{d}
\sum_{n_0,w_2 \in \mathbb{Z}} 
\rme^{-2\pi t\left(\frac{2 \pi n_0^2}{\sigma L_0^2} 
	+ \frac{\sigma(R^2 + w_2^2 L^2)}{2\pi}\right)}\nonumber\\
& = \left(\frac{1}{\eta(q)}\right)^{d}
\sqrt{\frac{\sigma}{4\pi t}}L_0
\sum_{m_0,w_2 \in \mathbb{Z}}
\rme^{- \frac{\sigma L_0^2}{4t} m_0^2 - t \sigma (R^2 + w_2^2 L^2)}~, 
\end{align}
where in the second step we performed a Poisson resummation, see eq. (\ref{poisson}).
As in the closed string case, the ghost partition function cancels exactly the non--zero--mode contributions of two bosonic fields: $Z_\tmb{gh}(t) = \eta^2(q)$.

To recover in our first-order string partition function (\ref{spfcyl}) the quantity $\hat\cZ_\tmb{cyl}$ that, according to eq. (\ref{prp0conazeff}), describes the Polyakov loop correlator, we have to select $m_0=1$, so that the target space cylinder $\cM$ is covered exactly once. Altogether we get 
\begin{align}
\label{eq:A.25}
\hat \cZ_\tmb{cyl} 
& = \sqrt{\frac{\sigma}{4\pi}}L_0  
\int_{0}^{\infty}  \frac{ dt }{2t^{\frac{3}{2}}}   
\left(\frac{1}{\eta(q)}\right)^{d-2}
\sum_{w_2 \in \mathbb{Z}} \rme^{- \frac{\sigma L_0^2}{4t} - t \sigma (R^2 + w_2^2 L^2)}\nonumber\\
&  = \sqrt{\frac{\sigma}{4\pi}}L_0 
\sum_{k=0}^\infty c_k  \sum_{w_2 \in \mathbb{Z}} 
\int_{0}^{\infty}  \frac{ dt }{2t^{\frac{3}{2}}}   
\rme^{- \frac{\sigma L_0^2}{4t} - t \big[ \sigma (R^2 + w_2^2 L^2  +  2\pi (k- \frac{ d-2 }{24}) \big]} ~.
\end{align}
where in the second line we expanded in series $\eta(q)$ according to eq. (\ref{expeta}).

We can perform the integral over $t$ using eq. (\ref{g12}), 
obtaining the simple expression
\begin{align}
\label{eq:cyl-open}
\hat \cZ_\tmb{cyl} 
& = \sum_{k=0}^\infty c_k  \sum_{w_2 \in \mathbb{Z}} 
\rme^{-L_0 \cE_\tmb{op}} ~.
\end{align}
where
\begin{align}
\label{cEcyOpenl}
\cE_\tmb{op} = \sigma \tilde R \sqrt{1 + \frac{2\pi}{\sigma   \tilde R^2} (k- \frac{ d-2 }{24})} ~.
\end{align}
and
\begin{align}
\label{tildeR}
\tilde{R} = \sqrt{ R^2 + w^2_2 {L}^2 }~.
\end{align}

This corresponds to the result of \cite{Billo:2005iv} in which the Polyakov loops distance $R$ is replaced by $\tilde{R}$, accounting for the effects of the further compactification in the transverse direction $x^2$.

\paragraph{Closed string channel}
On the other hand, the cylindrical world-sheet $\Sigma$ can also be viewed, interchanging the r\^ole of the w.s. coordinates $\xi^0$ and $\xi^1$, as the tree level propagation of a closed string between two boundary states representing the Polyakov loops. 
This interpretation is made explicit  rewriting the first line of eq. (\ref{eq:A.25}) 
in terms of the variable $s= 1/t$.
Taking into account that under this transformation we have
\begin{align}
\label{trasfnetat}
\eta\left(\rme^{-2\pi/s}\right) = \sqrt{s}  \eta\left(\rme^{-2\pi s}\right)~,
\end{align}
see eq. (\ref{Seta}), and expanding now in series $\eta\left(\rme^{-2\pi s}\right)$ we get
\begin{equation}
\label{eq:A.27}
\mathcal{Z}_\tmb{cyl} = \sqrt{\frac{\sigma}{4\pi}}L_0   
\sum_{k =0}^\infty c_k \sum_{w_2 \in \mathbb{Z} }
\int_{0}^{\infty} \frac{ds}{2s} s^{ \frac{3-d}{2}} 
\rme^{ - s \left(\frac{\sigma L_0^2}{4}  + 2\pi (k- \frac{ d-2 }{24})\right)  
	- \frac{\sigma}{s}(R^2 + w_2^2 {L}^2 )}~.
\end{equation}

Performing the integral over $s$ with the help of eq. (\ref{eq:A.16.2}) one gets
\begin{equation}
\label{eq:4.6}
\hat{\mathcal{Z}}_\tmb{cyl}   
= \sqrt{\frac{\sigma}{4\pi}}L_0 
\sum_{k=0}^{\infty} c_k \sum_{w_2\in\mathbb{Z}}
\left(\frac{\cE_\tmb{cl}}{2\tilde v}\right)^{\frac{d-3}{2}} 
K_{\frac{d-3}{2}}(\sigma\tilde\cA\cE_\tmb{cl})~, 
\end{equation}
where
\begin{equation}
\label{eq:4.7}
\tilde{\cA} = L_0 \tilde R~,~~~ \tilde v = \tilde R/L_0~,
\end{equation}
with $\tilde R$ given by eq. (\ref{tildeR}), and
\begin{align}
\label{cEcyl}
\cE_\tmb{cl} = \sqrt{1 + \frac{8\pi }{\sigma L_0^2} \left(k - \frac{d-2}{24}\right)}~.
\end{align}

In the $L \to \infty$ limit, due to the exponential asymptotic behaviour of the modified Bessel functions, the sum over $w_2$ is dominated by the term $w_2=0$  and all the higher values of $w_2$ are exponentially depressed. We recover again the result of \cite{Billo:2005iv}:
\begin{align}
\label{Zpolnc} 
\hat{\cZ}_\tmb{cyl} 
=\sqrt{\frac{\sigma}{4\pi}} L_0
\sum_{k=0}^{\infty} c_k
\left(\frac{\cE_\tmb{cl}}{2 u}\right)^{\frac{d-3}{2}} 
K_{\frac{d-3}{2}}(\sigma\cA\cE_\tmb{cl})~, 
\end{align} 
with $\cE$ still given by eq. (\ref{cEcyl}) and $u = R/L_0$.

As we mentioned, in the cylinder set-up the closed string propagates between two boundary states that represent the Polyakov loops in the closed string Hilbert space. In the stringy language the Polyakov loops are D0-branes, see the remark after eq. (\ref{eq:A.21}). Namely, they are $(0+1)$--dimensional objects, extended in the direction $x^0$, on which open strings can end. In presence of D0-branes, the $T$-duality
$L \to 2\pi/(L\sigma)$ does not map the theory to itself; rather it maps it to a theory in which the $D0$-branes become $D1$-branes, i.e. 2-dimensional objects extended also in the direction $x^2$. When we lower $L$ below $L_{c}$, we can re-express the cylinder partition function in terms of the theory with $D1$-branes and with the dual compactification  length $2\pi/(L\sigma)$, or we can remain within the $D0$--brane description. The length scale $L_{c}$ looses the meaning of a minimum compactification length. This fact has deep consequences in the LGT framework since it allows to decrease ``ad libitum'' the compactification radius thus allowing to study in great detail, as we shall see in the next section, the properties of the chromoelectric flux tube.

\subsection{Relation to the space--like string tension in LGT}
\label{subsec:LGT}
In the Lattice Gauge Theory context, the results discussed in the previous section have a natural application in the study of the so called ``space--like'' string tension 
\cite{Karkkainen:1993ch,Bali:1993tz,Karsch:1994af,Caselle1994f,Caselle:1993cb,Koch:1994zt,Ejiri:1995gd,Sekiguchi:2016gxx,Schroder:2005zd,Liddle:2007uy,
	Cheng:2008bs,Maezawa:2007fc,Alanen:2009ej,Andreev:2006eh,Andreev:2007rx,Meyer:2005px}. 
In this framework the compact transverse direction is identified with the inverse temperature of the theory. In this setting the Polyakov loops whose correlator we are studying lay in a space--like plane of the lattice and do not represent the order parameter of deconfinement. 

As we mentioned in the introduction the string tension extracted from such correlators is different from the actual string tension of the model (which is instead extracted from time--like Polyakov loop correlators) and  is denoted as ``space--like'' string tension to avoid confusion.  At low temperature, i.e. for large values of $L$ in our setting, the two string tensions coincide but as the temperature increases they behave in a 
different way \cite{Karkkainen:1993ch,Bali:1993tz,Karsch:1994af,Caselle1994f,Caselle:1993cb}. The ordinary string tension decreases as the deconfinement temperature is approached and vanishes at the deconfinement point, while the space--like one  remains constant and instead of vanishing it increases in the deconfined phase \cite{Karkkainen:1993ch,Bali:1993tz,Karsch:1994af}.  The physical reason for this behavior is that the correlator of two space--like Polyakov loops describes quarks moving in a finite temperature environment and it can be shown that what we call ``space-like string tension'' is actually related to the screening masses in hot QCD \cite{Koch:1994zt,Ejiri:1995gd,Sekiguchi:2016gxx,Schroder:2005zd,Liddle:2007uy,Cheng:2008bs,Maezawa:2007fc}, and thus it does not vanish in the deconfined phase. The precise temperature $T_\tmb{c}$
at which this change of behaviour of the space--like string tension occurs is not known. It is near the deconfinement point but there is no physical reason to expect that it should coincide with it.  Moreover it is not clear from the simulations if this change of behaviour indicates an actual critical point or simply a crossover.

The effective string theory description discussed above can be used to shed light on these issues. To this end however it is mandatory to take in a meaningful way the continuum limit, i.e. to
relate $\sigma$ and the finite temperature $1/L$ to well defined physical observables in the continuum limit. In the present setting this is much less obvious than for the ordinary string tension.
In the standard case, for time--like Polyakov loop correlators, it is easy to identify the deconfinement phase transition because the Polyakov loop is itself an order parameter. This allows to set the scale of the model, typically by measuring $T_\tmb{c}$ in units of the zero temperature string tension $\sigma$, and from this to set the temperature of the model by simply measuring it%
\footnote{
All these steps are usually performed using numerical estimates of $T_\tmb{c}$ and $\sigma$ extracted from Montecarlo simulations, but it is interesting to notice that the same analysis can be performed within the  framework of the effective string description, without resorting to any Montecarlo simulation and nevertheless with an impressive agreement with the numerical results, 
by  using the Olesen relation \cite{Olesen:1985e} . This relation
essentially amounts to identify the finite temperature deconfinement transition of the LGT with the Hagedorn transition of the effective string model \cite{Caselle:2015tza}, which relates $T_\tmb{c}$ and $\sigma$ as $T_\tmb{c}=\sqrt{\frac{3\sigma}{(d-2)\pi}}$.} in units of $T_c$. In our case the situation is the other way around, we have no insight on the location of the (putative) phase transition and there is in principle no obvious way to relate the inverse temperature $L$ with the space--like string tension $\sigma$ and thus give a physical scale in the continuum to measure and relate temperatures and masses. 
We shall see in the next section that describing the model in terms of a $T\bar T$  perturbation allows to make relevant progress in this direction. 

\section{$T\bar T$ perturbation of a compactified boson.}
\label{sec:TTbar}
In this section we re-consider the previous computation
from a diffrent perspective. As we mentioned in the introduction -- and as we will recall very briefly below -- the NG string in a $d$-dimensional target space represents the $T\bar T$ perturbation of the theory of $d-2$ free bosons. The spectrum of a $T\bar T$--deformed theory can be obtained from the unperturbed spectrum through a differential equation of the Burger's type. Also the explicit expression of the partition function satisfies differential constraints and could in principle be determined, but this is in practice not so trivial. In particular, the $T\bar T$ deformed partition function of compactified bosons was not yet written down. But this is exactly what we obtained, using the NG formulation, in the previous section. In fact we will show that the NG theory with one compact transverse direction can be described as the the $T\bar T$ deformation of the free bosonic theory, with one field being compact.  Remarkably, we shall also show that the $T\bar T$ deformation induces well defined trajectories in the parameter space of the model and can be performed only imposing a constant value of a suitable dimensionless combination of the compactification radius and the string tension of the model.  

\subsection{The NG theory as a $T\bar T$ perturbation}
\label{subsec:relTTbar}
We noticed in section \ref{subsec:NG} that the next to leading terms in the derivative expansion of the NG action in the physical gauge, eq.(\ref{sa}),
can be understood as a perturbation of the free bosonic action.
It can be shown \cite{Caselle:2013dra} that this is actually an integrable perturbation and that it coincides with the $T\bar T$ perturbation of the free action $S_0[X]$. In fact, the Nambu-Goto action can be rewritten as 
\begin{align}
\label{SN2}
S_\tmb{NG}=S_\tmb{cl} + S_{0}[X] - \frac1{2\sigma} \int d^2\xi\, T \bar{T}~.
\end{align}
The perturbing operator which appears in the above equation is the energy momentum tensor of the deformed theory and thus its explicit form must  be evaluated order by order in the $1/\sigma$ expansion. 

The normalization of the $T \bar{T}$ term above, by comparison with eq. (\ref{SN}), relates the $T\bar T$ perturbation parameter $t$ to the string tension $\sigma$ as follows:
\begin{align}
\label{ttosigma}
t = \frac{1}{2\sigma}~.
\end{align}
This normalization can be determined \cite{Caselle:2013dra} considering the first order term in the expansion, namely the one in eq. (\ref{s1pg}). In this case the energy-momentum tensor of the free-field theory (\ref{gauss}) is simply given by
\begin{align}
\label{Tab}
T_{\alpha\beta}=\partial_\alpha X\cdot\partial_\beta X-\frac12\delta_{\alpha\beta}
\left(\partial^\gamma X\cdot\partial_\gamma X\right)~.
\end{align}
It is easy to see that, up to the  the next to leading term, eq. (\ref{sa}) can be rewritten as
\begin{align}
\label{sat}
S_\tmb{NG} = S_\tmb{cl} + S_0[X]-\frac\sigma4 \int d^2\xi\, T_{\alpha\beta}T^{\alpha \beta}+\dots~.
\end{align}
If we now rewrite the perturbing term using chiral components\footnote{Notice that this is not the standard normalization of the energy momentum tensors which would instead require an additional factor $\pi$: 
	$T = - \pi \sigma (T_{11}-iT_{12})~,~~~ \bar T = - \pi \sigma (T_{11}+iT_{12})$
	so as to obey the standard OPE relation $ T(z)T(w)=\frac{d-2}2\frac1{(z-w)^4} + ...~$ with $z = \xi^0 +\ii \xi^1$. We chose the normalization of eq.(\ref{TTbardef}) to conform with the standard notations of the $T\bar T$ literature.}
\begin{align}
\label{TTbardef}
T = -\sigma (T_{11}-iT_{12})~,~~~ 
\bar T = - \sigma (T_{11}+iT_{12})
\end{align}
we end up with the normalization appearing in  
eq.(\ref{SN2}).

The above analysis can be extended to all order in the perturbing parameter \cite{Smirnov:2016lqw, Cavaglia:2016oda} and induces a set of constraints 
on the spectrum of the theory and in particular on the dependence of the energy levels 
on the perturbing parameter $t$.
Let us assume that the theory is defined on a two-dimensional manifold with a compact space-like direction of size $L_1$. The spatial momentum $P(L_1)$ of any state is quantized in unities of $2\pi/L_1$  and is preserved along the perturbation. The energy $E(L_1,t)$ of the state depends in general both on $L_1$ and on the perturbing parameter $t$. Remarkably enough the constraints alluded to above can be summarized in  
the requirement that the energy levels
satisfy the following inhomogeneous Burgers equation \cite{Smirnov:2016lqw, Cavaglia:2016oda}
\eq
\label{Burgers}
\frac{\partial E}{\partial t} = \frac{1}{2} \frac{\partial\left( E^2 - P^2 \right)}{\partial L_1}~.
\en

\paragraph{The CFT of a free non-compact boson}
It is useful for the following analysis to look at the solution of this equation in the case of a 2d free bosonic theory (see \cite{Smirnov:2016lqw, Cavaglia:2016oda,Conti:2018jho} for a detailed derivation).
A quantum state of the unperturbed theory is characterized by the left- and right-moving oscillation numbers $k$ and $k^\prime$; its unperturbed energy and momentum are given by
\eq
\label{Burgers1}
E(0,L_1) = \frac{2\pi}{L_1} \left(k + k^\prime - \frac{1}{12}\right)
\equiv \frac{2\pi}{L_1} \varepsilon~,~~~
P(L_1) = \frac{2\pi}{L_1}(k - k^\prime) 
\equiv \frac{2\pi}{L_1} p~.
\en
The subscript $(0)$ means ``unperturbed value'' and we introduced $\varepsilon = k + k^\prime - 1/12$ and $p = k - k^\prime$ to avoid typographical clutter in the following formul\ae\,.  
The general solution to (\ref{Burgers}) with these initial conditions is
\eq
\label{Burgers2}
E(t,L_1)  = \frac{L_1}{2 t} \left( -1 + \sqrt{1 + \frac{8\pi t}{L_1^2} \varepsilon + \frac{16\pi^2 t^2}{L_1^4} p^2} \right)  ~.
\en

In our case however there is an additional subtlety.
Since we are interested in particular in mapping the perturbed spectrum onto the Nambu Goto one we must also keep into account the possible  presence of an additional ``classical'' energy term -- the term denoted as $S_\tmb{cl}$ in eq.(\ref{SN2}). One has to  start in this case with the unperturbed spectrum 
\eq
E(0,L_1) = \frac{2\pi}{L_1}\varepsilon + F_0 L_1 \;,
\label{Burgers3}
\en
and solving the Burgers equation following \cite{Smirnov:2016lqw,Conti:2018jho} one finds 
\eq
E(t,L_1) = {\tilde F L_1} + \frac{L_1}{2\tilde{t}} \left( -1 + \sqrt{1 + \frac{8\pi\tilde{t}}{L_1^2}\varepsilon + \frac{16\pi^2\tilde{t}^2}{L_1^4}p^2} \right) \;,
\label{Burgers4}
\en
where
\begin{equation}
\label{tfis}
\tilde{t} = t(1-t F_0)~,~~~	\tilde F=\frac{F_0}{1-t\,F_0}~.
\end{equation} 
If we assume%
\footnote{
	Notice that as a consequence of these assumptions the string tension of the unperturbed model does not coincide any more with the Nambu-Goto one \cite{Nando2019}; in fact, one has $F_0 = \sigma/2$ and $t = 1/\sigma$. This fact
	has no direct relevance for the present analysis, but it might be important when using  these $T\bar T$ deformed models to describe the confining regime of
	Lattice Gauge Theories \cite{Nando2019}. We plan to address this issue in a future work.}
$\tilde{t}=1/(2\sigma)$ and $\tilde F=\sigma$, the resulting energy spectrum the Nambu-Goto string model, namely
\eq
\label{Burgers5}
E(t,L_1) = \sigma L_1 \cE(t,L_1) =  \sigma L_1 
\sqrt{1 + \frac{4\pi}{\sigma L_1}\varepsilon + \frac{4\pi^2}{(\sigma L_1)^2} p^2}~.
\en
coincides with the one obtained for the Nambu-Goto string model in $d=3$ with a single transverse direction taken to be non-compact, see eq.s (\ref{eq:1.42},\ref{torus3}).

These results, and the constraints imposed by the Burgers equation will play a major role in the following analysis, in which we consider the presence of a compactified transverse dimension in the model.
We will first discuss the CFT of a free compactified boson which represents the unperturbed CFT in such a setting. We shall then obtain its $T\bar T$ deformation in sect.\ref{subsec:TTpert} from the NG partition function derived earlier. 

\subsection{The CFT of a free compactified boson}
\label{sec:comp}
Let us consider the free bosonic theory with the action in eq. (\ref{gauss}), defined on the 2-dimensional manifold $\cM_0$ which represents, in the effective string perpective, the minimal surface swept out in target space by the string. Let us now assume that a bosonic field $X$, which in the set-up of section \ref{sec:EST} we took to be $X_2$, obeys the compactification condition 
\begin{equation}
\label{compX}
X \sim X + L~.
\end{equation}
This changes drastically the  behaviour of the  2d theory of the field $X$. 
In particular, a marginal operator  appears in the spectrum and, as a consequence, a whole line of critical points. The action is given by eq. (\ref{gauss}), restricted to the single field $X$. The coupling constant%
\footnote{Often in the literature the coupling constant is defined as $g=2\pi\sigma$.} $\sigma$ allows to tune the model along the critical line and to introduce 
the dimensionless compactification radius 
\begin{equation}
\label{defrho}
\rho=\sqrt{\frac{\sigma}{2\pi}} L~.
\end{equation}
The CFT of a compactified boson is characterized by a rich spectrum of 
primary operators $O_{n,w}$ whose scaling dimensions are labeled by two indices $n$ and $w$ (which in the XY model label the ``spin'' and ``vortex'' sectors respectively) 
and depend on  $\rho$ as follows%
\footnote{\label{foot-ginsparg}See for instance \cite{Ginsparg:1988ui} for further details, but note that the compactification radius $r$ of \cite{Ginsparg:1988ui} is equivalent to our $\rho/\sqrt{2}$.}:
\eq
\label{CFT1}
h_{n,w}(\rho)= \frac12\left(\frac{n^2}{\rho^2}+ w^2\rho^2\right)~.
\en
These weights exhibit a $\rho \to 1/\rho$ ``duality'' symmetry which exchanges spin and vortex sectors. In the String Theory language, this is known as $T$-duality, see eq. (\ref{Tduality}). 

This CFT is well defined for any value of $\rho$ but for some choices of $\rho$ additional symmetries emerge which make the theory particularly interesting. In particular for 
$\rho=1$ -- i.e., at the self--dual point -- one has the level 1 SU(2) WZW model.  For $\rho=\sqrt{2}$ one finds the CFT of a free Dirac fermion and finally for $\rho=2$ one has the famous Kosterliz-Thouless critical point. There are several lattice realization of this CFT. The most important ones are the XY model, which is defined for $\rho>2$ (or equivalently $\rho<1/2$) and the SOS model, which can be shown to be equivalent to the well known six vertex model and which is defined for all values of $\rho$.

The toroidal partition function corresponds to the euclidean path integral of this model, with the action (\ref{gauss}), when the base manifold $\cM_0$ is a torus; this is the situation considered in the effective string description of the interface free energy, see section \ref{subapp:int}.
Let us denote by $\tau_0$ the complex structure modulus of $\cM_0$; in the case considered in section \ref{subsec:effsetup}, see eq. (\ref{AuM0}), we simply have 
\begin{align}
\label{tau0}
\tau_0 = \ii\, u = \ii L_0/L_1
\end{align} 
as we consider a straight torus, but the following formul\ae\, hold also when the modulus $\tau_0$ has a real part as well. We reviewed the computation of the partition function for the compact boson \cite{Ginsparg:1988ui}
in section \ref{subapp:int}, with the result given  eq. (\ref{CFT2s}), which we rewrite here in terms of the adimensional compactification parameter $\rho$ and of the modulus $\tau_0$ through the quantity
\begin{align}
\label{defq0}
q_0 = \exp(2\pi\ii\tau_0) = \exp(-2\pi u)~.
\end{align}
Thus we have%
\footnote{We denote this partition function as $Z_{(0)}$ to remark that it corresponds to the free case in absence of $T\bar T$ deformation.}
\begin{align} 
\label{CFT2}
Z_{(0)}(\rho) & = \frac{ 1 }{ \eta(q_0)}\frac{ 1 }{ \eta( \bar q_0)} 
\sum_{n, w \in \mathbb{Z}} q_0^{  \frac{1}{4} \big( \frac{n}{\rho} + {w\rho} \big)^2 } \bar{q_0}^{  \frac{1}{4} \big( \frac{n}{\rho} - {w\rho} \big)^2 }
\notag\\
& = \sum_{n,w\in\mathbb{Z}} \sum_{k,k^\prime= 0}^\infty p_k p_{k^\prime}
\rme^{- 2\pi u \left(\frac 12 \left(\frac{n^2}{\rho^2} + w^2 \rho^2\right) + k + k^\prime - \frac{1}{12}\right)}~,
\end{align}
where in the second step we have used eq. (\ref{expeta}). Writing for each term in the sum the exponent as 
$-L_0 E(0,L_1)$ this corresponds to the energy spectrum
\begin{align}
\label{Efreecomp}
E(0,L_1) =  \frac{2\pi}{L_1} \left(h_{n,w}(\rho) + k + k^\prime - \frac{1}{12}\right)
\equiv \frac{2\pi}{L_1} \tilde\varepsilon
\end{align} 
where, in each sector labeled by $n$ and $w$ we have, on top of the contribution of eq. (\ref{CFT1}), that of right-moving and left-moving oscillation modes.
We use the notation $E(0,L_1)$ to stress that this spectrum is at zero $T\bar T$ perturbation. 
Let us note that these states also have momentum 
\begin{align}
\label{Pfreecomp}
P(L_1) = \frac{2\pi}{L_1} (k - k^\prime + n w) \equiv \frac{2\pi}{L_1} \tilde p~.
\end{align} 
With respect to the non-compact free boson spectrum of eq. (\ref{Burgers1}) we simply replace $\varepsilon$ and $p$ with $\tilde{\varepsilon}$ and $\tilde p$ which depend also on the additional quantnotableum numbers $n,w$.  

\subsection{The partition function of the $T\bar T$ perturbed theory.}
\label{subsec:TTpert}
It is natural to expect the relation between $T\bar T$ perturbed models and the Nambu-Goto effective action to hold also in presence of a transverse compactified dimension. The Nambu-Goto model interface partition function computed in section \ref{sec:EST}, with $d=3$ and with the single transverse direction being compact, should
therefore represent the partition function of the $T\bar T$ deformation of a free compact boson. 
Let us  consider the result of eq. (\ref{eqA}) for $d=3$. Using eq. (\ref{K12}) and 
renaming for simplicity $n$ and $w$ the integers $n_2$ and $w_2$ appearing in eq. (\ref{eqA}) we get%
\footnote{The coefficients $c_k$ appearing in eq. \ref{eqA} reduce, for $d=3$, to the numbers $p_k$ of partitions of the integer $k$ appearing in eq. (\ref{CFT2}.}
\begin{align}
\label{eqA3d}
\hat\cZ_\tmb{torus} = \frac{1}{4}  
\sum_{n_1,n,w\in\mathbb{Z}} \sum_{k, k^\prime = 0}^\infty p_k p_{k^\prime}
\delta_{n_1 + k-k^\prime + n w}
\rme^{-\sigma L_0 L_1 \cE}~,
\end{align} 
where 
\begin{align}
\label{cEd3c}
\cE = \sqrt{1 + \frac{4\pi}{\sigma L_1^2}\left(k+k^\prime - \frac{1}{12}\right) 
	+ \frac{4\pi^2 n_1^2}{\sigma^2 L_1^4}
	+ \frac{4\pi^2 n^2}{\sigma^2 L_1^2 L^2} 
	+ \frac{w^2 L^2}{L_1^2}}~.
\end{align}
To see that this expression corresponds to energy levels $E = \sigma L_1 \cE$ which satisfy the Burgers' equation 
we have to rewrite it in terms of the dimensionless compactification parameter $\rho$ introduced in eq. (\ref{defrho}), setting
\begin{align}
\label{Ltorho}
L = \sqrt{\frac{2\pi}{\sigma}} \rho~.
\end{align}
Moreover, the Kronecker delta in eq. (\ref{eqA3d}) identifies (minus) the integer $n_1$ with the momentum $\tilde{p}$ appearing in eq. (\ref{Pfreecomp}):
\begin{equation}
\label{n1is}
- n_1 = k - k^\prime + n w = \tilde p~. 
\end{equation}
In this way we obtain
\begin{align}
\label{Ettcomp}
\sigma L_1 \cE & = \sigma L_1 \sqrt{ 1 
	+ \frac{4\pi}{\sigma L_1^2}\left[\frac 12 \left(\frac{n^2}{\rho^2} + w^2 \rho^2\right) + k + k^\prime - \frac{1}{12}\right] + \frac{4\pi^2 n_1^2}{\sigma^2L_1^4}}\nonumber\\
& =  \sigma L_1 \sqrt{1 + \frac{4\pi}{\sigma L_1^2}\tilde{\varepsilon} 
	+  \frac{4\pi^2}{\sigma^2L_1^4}{\tilde p}^2}~,
\end{align}
where $\tilde\varepsilon$ was defined in eq. (\ref{Efreecomp}) and $\tilde p$ in eq. (\ref{Pfreecomp}). 
We see that this expression takes the form of eq. (\ref{Burgers5}), namely that of the energy levels $E(t,L_1)$
of the $T\bar T$ perturbation of a theory that at $t=0$ has the energy levels and momenta characterized by the  the energy $\tilde  \varepsilon$ and the momentum $\tilde p$. This unperturbed theory is exactly the free compactified boson.

In other words, the Nambu-Goto model with one compact transverse dimension can indeed be interpreted as the $T\bar T$ deformation of the free compactified boson, and the torus NG partition function (\ref{eqA3d}) provides the partition function for this deformed theory. Let us remark however that
it is crucial to use eq. (\ref{Ltorho}) before mapping the string tension $\sigma$ to the perturbing parameter $t$. In the 2d space of couplings $(\sigma,L)$ the $T\bar T$ perturbation defines trajectories at fixed $\rho$ and thus sets a well defined relation -- modulated by the marginal parameter $\rho$ -- between the compactification scale $L_2$ and the perturbation parameter $\sigma$. This is exactly the relation that we were looking for and that will allow us in the next section to take a sensible continuum limit of our lattice observables. 

\section{Implications for LGTs.}
\label{sec:conclusion}
Let us now come back to the result of section \ref{subapp:pol} on the expectation value the Polyakov loops correlator in presence of a compactified transverse dimension. Within the Nambu-Goto effective string description this expectation value is represented by the cylinder partition function:
\begin{align}
\label{PPisZcyl}
\vvev{P^\dagger(R)P(0)} \equiv \hat{\cZ}_\tmb{cyl}
\end{align}
and we  have obtained exact expressions of $\hat{\cZ}_\tmb{cyl}$ both in the open string channel, eq. (\ref{eq:cyl-open}), and in the closed string channel, eq. (\ref{eq:4.6}). These exact results could be compared with numerical simulations of this observable in Lattice Gauge Theories to investigate the extent of validity of the effective Nambu-Goto description. In the following, we will focus on the $d=3$ case for simplicity.

In particular, if we consider the closed string channel expression, and assume a large distance $R$ between the two Polyakov loops, the exact result simplifies and takes a form which displays clearly the effects of the compactification. For large $R$ the amplitude is dominated by the exchange of the lowest-lying closed string state, the so-called tachyon. To be more explicit let us introduce, extending what we did for the transverse compactification scale in eq. (\ref{defrho}), the dimensionless quantities
\begin{align}
\label{defrl0}
r = \sqrt{\frac{\sigma}{2\pi}} R~,~~~
\rho = \sqrt{\frac{\sigma}{2\pi}} L~,~~~
l_0 = \sqrt{\frac{\sigma}{2\pi}} L_0~.
\end{align}

\rev{In terms of these quantities we can single out different interesting regimes. For instance, choosing 
\begin{align}
\label{regime}
r \sim l_0 \gg \rho~,
\end{align}
the effect of the compactification radius is emphasized. Testing in this regime our predictions  against 
the $\rho$ dependence  of lattice results one could investigate the possible presence of a critical value $\rho_c$ (see below).

Another interesting limit is the one in which
\begin{align}
\label{regime2}
r  \gg \rho \gg l_0~.
\end{align}
In this limit, the smallest length scale is the length of the Polyakov loops and, as we mentioned in  sect. \ref{subsec:NG},  this is the situation in which one expects to find the largest deviations of lattice data with respect to the NG predictions. Thus, starting from a precise prediction of our first-order NG model for this compactified situation, we may hope to use the comparison with the corresponding lattice results to gain further insight into the higher order corrections to the confining string action.
To this end it is mandatory  to isolate precisely the corrections due to the compactification radius $\rho$ from the remaining terms.  We shall show below that this is indeed possible and thus that this particular geometry is perfectly suited to identify subtle correction terms in the confining string action.}

\rev{In the $r  \gg \rho \gg l_0$ limit}, the arguments of the Bessel functions appearing in eq. (\ref{eq:4.6}) are large. Thus the Bessel functions themselves have a decaying exponential behaviour, see eq. (\ref{asK}), and the terms with $k >0$ are exponentially suppressed with respect to the term with $k=0$. Note that also increasing $w$ yields contributions which are suppressed, but less steeply than increasing $k$. We will thus 
consider the lowest values of $|w|$ to account for the effect of a large but finite compactificaton parameter $\rho$.

We remain with
\begin{align}
\label{k0only}
\vvev{P^\dagger(R)P(0)} 
= \frac{l_0}{\sqrt{2}} \sum_{w\in\mathbb{Z}} K_0(\sigma \tilde{\cA} \cE_0)
\sim \frac{l_0}{\sqrt{2}} \sum_{w\in\mathbb{Z}} \sqrt{\frac{\pi}{2\sigma \tilde{\cA} \cE_0}}
\rme^{-\sigma \tilde{\cA}\cE_0}~.	
\end{align}
Here $\cE_0$ corresponds to the expression in eq. (\ref{cEcyl}) with $d=3$ and $k=0$, namely to
\begin{align}
\label{CE0}
\cE_0 = \sqrt{1 - \frac{1}{6 l_0^2}}~.
\end{align}
Note that we have to assume $l_0 > 1/\sqrt{6}$, that is $L_0 > \sqrt{\pi/3\sigma}$ to avoid the tachyon singularity in which $\cE_0$ vanishes.
Moreover from eq.s (\ref{eq:4.7},\ref{tildeR}) and (\ref{defrl0}) we have
\begin{align}
\label{sAt}
\sigma \tilde{\cA} = 2	\pi l_0 r \sqrt{1 + \frac{\rho^2}{r^2} w^2} 
\sim 2	\pi l_0 r \left(1 + \frac{w^2}{2} \frac{\rho^2}{r^2} + \ldots\right)~,
\end{align}
where in the second step we took into account eq. (\ref{regime}). 

Let us keep in eq. (\ref{k0only}) the contributions of the lowest winding numbers, $w=0$ and $w=\pm 1$. With a straightforward expansion in the small ratio $\rho/r$ we get
\begin{align}
\label{new5}
\vvev{P^\dagger(R)P(0)} 
\sim \sqrt{\frac{l_0}{8r\cE_0}}
\rme^{-2\pi l_0r\cE_0} \left(1 + 2 \left(1 - \frac 14 \frac{\rho^2}{r^2}\right)
\rme^{-\pi l_0 \cE_0 \frac{\rho{^2}}{r}}\right)~.  
\end{align}
We see that the correction associated to the transverse dimension has a very peculiar $1/R$ dependence
which should make it accessible to numerical simulations.
It should be possible, either using the exact expressions or the approximation we just described, to compare efficiently numerical simulations of the Polyakov correlator in Lattice Gauge Theories with a compact transverse direction with the Nambu-Goto prediction; we plan to do so in a future work. 

\rev{Let us go back to the regime introduced in eq. (\ref{regime}).} One expects, decreasing the size $\rho$ of the compact direction
to reach a critical value $\rho_\tmb{c}$ where the NG predictions cease to describe correctly the data. A reliable identification of $\rho_\tmb{c}$ would represent an important piece of evidence with respect to some existing conjectures about the behaviour of this observable in Lattice Gauge Theories. Let us discuss a couple of specific issues. 

It was proposed a few years ago by Meyer \cite{Meyer:2005px} that, at the critical value of the compact direction, the theory undergoes a dimensional reduction
which could be described as a Kosterlitz-Thouless (KT) phase transition. The idea behind this conjecture is to consider the topologically non trivial windings of the flux tube around the compactified transverse dimension as vortices in the world sheet of the effective string which condense in the vacuum at  the point at which dimensional reduction occurs. 
In the normalizations that we introduced in section \ref{sec:comp} for the $c=1$ CFT of a compact boson, see eq. (\ref{CFT1}) and footnote \ref{foot-ginsparg}, the KT point corresponds  to $\rho=2$.
If the $\rho_\tmb{c}$ value at which our NG result for the Polyakov loop correlators ceases to agree with the lattice data turns out to be close to $2$, it would support this conjecture.  As a side remark let us notice that our analysis suggests that, if the conjecture holds, we should better expect the dimensional reduction point to be represented by the $T\bar T$ deformation of a KT point. Such a critical point would represent the first physical realization of a $T\bar T$ deformed Kosterlitz-Thouless transition.

Another important aspect that could be better understood following a precise determination of $\rho_\tmb{c}$ is related to the effects of the intrinsic width%
\footnote{For an introduction to the intrinsic width of the flux tube in the confining regime of lattice gauge theories see for instance \cite{Caselle2012a} and the references therein.}
of the flux tube. 
One expects that the effective string picture holds only for values of the transverse dimension larger than this scale.
This is indeed the main difference between the real flux tube of a confining gauge theory and the Nambu-Goto string which instead has a (classically) vanishing thickness%
\footnote{The quantity which is usually called in the LGT context ``width of the Nambu-Goto string'', which diverges logarithmically with the interquark distance, is only due to quantum fluctuation \cite{Luscher:1980iy} and should not be confused with the \emph{intrinsic} width that we discuss here. 
	The actual width of the flux tube is the sum of the two.}

The correlator of space--like Polyakov loops is a perfect tool to identify and measure this intrinsic width. In fact by increasing enough the temperature, i.e., decreasing the transverse size $L$, at some point it will reach the intrinsic width of the flux tube. This point can be easily recognized because, once this threshold is reached, if one keep increasing the temperature,i.e., if one starts to ``squeeze'' the flux tube below its intrinsic width,
the space--like string tension starts to increase\cite{Karkkainen:1993ch,Bali:1993tz,Karsch:1994af}. This phenomenon is understood in the effective string framework as due to the increase in the flux density within the flux tube due to its squeezing \cite{Karkkainen:1993ch,Bali:1993tz,Karsch:1994af,Caselle1994f,Caselle:1993cb}. 
With our analysis we may associate a value $\rho_\tmb{c}$ to this threshold and perform a well defined continuum limit for this quantity. 
It would be very interesting to see if this value is universal or if it depends on the particular gauge theory under study and if it is related to the KT transition mentioned above or to some other special point along the $\rho$ line.

\vskip 1.5cm
\noindent {\large {\bf Acknowledgments}}
\vskip 0.2cm
We thank R. Conti, F. Gliozzi and R. Tateo for 
many useful discussions and suggestions.
\noindent
The work of M.B. is partially supported by the MIUR PRIN Contract 
2015 MP2CX4 ``Non-perturbative Aspects Of Gauge Theories And Strings''.
\vskip 1cm
\begin{appendix}

\section{Details on the derivation of the partition functions}
\label{app:A}

Let us consider the zero-mode factor in the directions $0$ and $1$, and let us write it as 
\begin{equation}
\label{ZL0L1}
\Gamma(\tau,\bar{\tau};L_0)\Gamma(\tau,\bar{\tau};L_1)
= \frac{\sigma L_0 L_1}{2\pi\tau_2}	\Gamma^{(2)}(\tau,\bar{\tau};L_0,L_1)~.
\end{equation}    
Starting from the last expression in eq. (\ref{Gammazm}) and using the manipulations introduced in \cite{Dixon:1990pc} we can write
\begin{align}
\label{G2DKL}
\Gamma^{(2)}(\tau,\bar{\tau};L_0,L_1) = 
\sum_A \rme^{\sigma L_0 L_1 \det A}
\exp\left(- \frac{\sigma L_0^2}{2\tau_2} 
\left|(1,\ii L_1/L_0) A \begin{pmatrix}\tau \\ 1\end{pmatrix}
\right|^2
\right)~,
\end{align}
where $A$ is the two--by--two integer matrix
\begin{align}
\label{Amatis}
A = \begin{pmatrix} w^0 & m^0 \\ w^1 & m^1\end{pmatrix}~. 
\end{align}
A modular transformation $\tau\to\tau^\prime$ as in eq. (\ref{modaction}) is easily seen to be equivalent to 
\begin{align}
\label{Aprime}
A \to A^\prime = A \begin{pmatrix} a & b \\ c & d\end{pmatrix}~;
\end{align} 
this amounts to the transformation (\ref{modularmw}) on both $m^0,w^0$ and $m^1,w^1$. The space of the matrices $A$ is partitioned into orbits of this action, under which
\begin{align}
\label{detA}
\det A = w^0 m^1 - w^1 m^0
\end{align}
is invariant. Given the meaning of the $m$'s and $w$'s as wrapping numbers, $\det A$ is the number of times the string wraps the target space torus in the direction $0$ and $1$.
In eq. (\ref{pftorus}) $\Gamma^{(2)}$, multiplied by other modular invariant factors, is integrated over the fundamental cell $\mathscr{F}$. Equivalently, as argued in \cite{Dixon:1990pc}, we can integrate $\tau$ over the entire upper half plane but limit the sum over the matrices $A$ to one representative for each orbit. To reproduce the interface free energy we have to pick up the orbit with $\det A = 1$, a representative of which is simply $A = \mathbf{1}$.  We can also restrict the integration over $\tau$ to the fundamental cell with respect to the subgroup of $\mathrm{PSL}(2,\mathbb{Z})$ given  by the translations $\tau \to \tau + b$ only, namely the perodic strip $-1/2 \geq \tau_1 \geq 1/2$, and sum in $\Gamma^{(2)}$ over all matrices of the form  
\begin{equation}
\label{Apartial}
A = \begin{pmatrix} 1 & m^0 \\ 0 & 1\end{pmatrix}~.
\end{equation} 

\section{Useful formul\ae}
\label{app:A}
We list here miscellaneous definitions, properties and formul\ae\, that we use in the main text.

Dedekind's $\eta$-function is defined as
\begin{align}
\label{defeta}
\eta(q)=q^{1/24}\prod_{n=1}^\infty(1-q^n)~.
\end{align}
It can be expanded in powers of $q$ as follows:
\begin{align}
\label{expeta}
\left[\eta(q)\right]^{-1} = \sum_{k=0}^\infty p_k q^{k-\frac{1}{24}}~,  
\end{align}
whee $p_k$ is the number of partitions of the integer $k$. We will use the following generalization of this expansion:
\begin{align}
\label{expetad}
\left[\eta(q)\right]^{-(d-2)} = \sum_{k=0}^\infty c_k q^{k-\frac{d-2}{24}}~.  
\end{align}

Under a modular transformation $\tau \to \tau^\prime$ with
\begin{align}
\label{modactionbis}
\tau^\prime  = \frac{a\tau + b}{c\tau + d}~,~~~a,b,c,d\in\mathbb{Z}~,~~~ad-bc=1~,
\end{align}
one has
\begin{align}
\label{modt2}
\tau_2 \to \frac{\tau_2}{\left|c\tau + b\right|^2}~,
\end{align}
while the following quantities are invariant:
\begin{align}
\label{modinvs}
\frac{d^2\tau}{\tau_2^2}~,~~~
\sqrt{\tau_2} \eta(q) \eta(\bar q)~,
\end{align}
where $q = \exp(2\pi\ii\tau)$. Under the $S$ modular transformation $\tau\to -1/\tau$, in particular, we have
\begin{align}
\label{Seta}
\eta\left(\rme^{- 2\pi\ii/\tau}\right) = \sqrt{-\ii\tau} \eta\left(\rme^{2\pi\ii\tau}\right)~.
\end{align}

The Poisson resummation formula states that
\begin{align}
\label{poisson}
\sum_{n, w \in \mathbb{Z}}\exp\left(-\pi a n^2 + 2 \pi\ii b n\right)
= a^{-\frac 12} \sum_{m \in\mathbb{Z}} \exp\left(- \frac{\pi(m-b)^2}{a}\right)~.
\end{align}

In the main text we make use of the the following integral: 
\begin{equation} 
\label{eq:A.16.2}
\int_{0}^{\infty} \frac{ dt }{t^{1+\gamma}} \rme^{- \alpha^2 t - \frac{\beta^2}{t}} = 2\left(\frac{\left|\alpha\right|}{\left|\beta\right| }\right)^\gamma K_\gamma(2\left|\alpha\right|\left|\beta\right|)~.
\end{equation}
In the case $\gamma=1/2$ this reduces simply to
\begin{align}
\label{g12}
\int_{0}^{\infty} \frac{ dt }{t^{3/2}} \rme^{- \alpha^2 t - \frac{\beta^2}{t}} =
\frac{\sqrt{\pi}}{|\beta|} \rme^{-2 |\alpha||\beta|}~,
\end{align}
in accord with the relation
\begin{equation}
\label{K12}
K_{1/2}(x) = \frac 12 \sqrt{\frac{2\pi}{x}} \rme^{-x}~.		
\end{equation}

The asymptotic expansion of the functions $K_\alpha(z)$ is of the form
\begin{align}
\label{asK}
K_\alpha(z) \sim \sqrt{\frac{\pi}{2z}} \rme^{-z}\left(1 + \frac{4\alpha^2-1}{8z} + \ldots\right)~.
\end{align}

\end{appendix}


\providecommand{\href}[2]{#2}
\begingroup\raggedright

\endgroup

\end{document}